\documentclass[preprint,11pt, number, sort]{elsarticle}

\usepackage{soul}
\usepackage{graphicx}
\usepackage{amsthm, amssymb, amsmath}
\numberwithin{equation}{section}

\usepackage{epstopdf}
\usepackage{subfigure}
\usepackage{algcompatible}
\usepackage{comment}
\usepackage{tikz}
\usepackage{graphicx}

\usepackage[hmarginratio=1:1,top=32mm,columnsep=20pt]{geometry}
\usepackage{mathtools}

\usepackage[pagebackref=true, colorlinks=true, urlcolor=blue, pdfborder={0 0 0}]{hyperref}
\usepackage[noabbrev,capitalize]{cleveref}
\usepackage[notref,notcite,color]{showkeys}

\usepackage[displaymath]{lineno}

\makeatletter
\let\LN@align\align
\let\LN@endalign\endalign
\renewcommand{\align}{\linenomath\LN@align}
\renewcommand{\endalign}{\LN@endalign\endlinenomath}
\let\LN@gather\gather
\let\LN@endgather\endgather
\renewcommand{\gather}{\linenomath\LN@gather}
\renewcommand{\endgather}{\LN@endgather\endlinenomath}
\makeatother

\renewcommand{\vec}[1]{\mathbf{#1}}
\def\ie{\rm{{\it i.e.}}}
\def\eg{\rm{{\it e.g.}}}
\def\X{\mathcal{X}}
\def\R{\mathbb{R}}
\def\E{\mathbb{E}}
\def\N{\mathsf{N}}
\def\x{\vec x}
\def\me{\text{me}}
\def\md{\text{md}}

\def\tas{TA$^2$S$^2$\ }
\newcommand{\rd}{\mathrm{d}}

\theoremstyle{definition}

\theoremstyle{remark}

\newcommand{\fgp}{f_{\textsf{GP}}}
\newcommand{\Igp}{I_{\textsf{GP}}}
\newcommand{\red}{\color{black}}

\newcommand{\nc}{\normalcolor}

\graphicspath{{./img/}}

\journal{Unknown}
\begin{document}

\begin{frontmatter}

\title{\red History matching with probabilistic emulators and active learning\nc}

\author[caltech]{A. Garbuno-Inigo \corref{cor1}}
\ead{agarbuno@caltech.edu}
\author[ucl]{F. A. DiazDelaO }
\author[caltech]{K. M. Zuev}
\cortext[cor1]{Corresponding author.}
\address[caltech]{Department of Computing and Mathematical Sciences,
California Institute of Technology, Pasadena, CA 91125, USA.}
\address[ucl]{{\red Clinical Operational Research Unit, Department of Mathematics,
 University College London\\ London, WC1H 0BT, United Kingdom.}}

\begin{abstract}

The scientific understanding of real-world processes has \red dramatically \nc
improved over the years through computer simulations. Such simulators represent
complex mathematical models that are implemented as computer codes which are
often expensive. The validity of using a particular simulator to draw accurate
conclusions relies on the assumption that the computer code is correctly
calibrated. This calibration procedure is often pursued under extensive
experimentation and comparison with data from a real-world process. The problem
is that the data collection may be so expensive that only a handful of
experiments are feasible. History matching is a calibration technique that,
given a simulator, it iteratively discards regions of the input space using an
implausibility measure. When the simulator is computationally expensive, an
emulator is used to explore the input space. In this paper, a \red Gaussian
process \nc provides a complete probabilistic output that is incorporated into
the implausibility measure. The identification of regions of interest is
accomplished with recently developed annealing sampling techniques. Active
learning functions are incorporated into the history matching procedure to
refocus on the input space and improve the emulator. The efficiency of the
proposed framework is tested in well-known examples from the history matching
literature, as well as in a proposed testbed of functions of higher dimensions.

\end{abstract}


\end{frontmatter}

\section{Introduction}

\red Computer simulations have played an essential role in vastly improving our
understanding of real-world processes. \nc These computational models are
parameterised by a set of values which determine the behaviour of the simulator
and its ability to replicate the process under consideration. No matter how
sophisticated or efficient, a simulator must be well-calibrated to experimental
data if it is to be trusted. Thus, the validity of using a particular simulator
to draw accurate conclusions, relies on the assumption that the model has been
correctly calibrated. That is, the vector of input parameters is well known, and
there is confidence that these values are able to replicate the process the
simulator is modeling. This calibration procedure is often pursued under
extensive code experimentation guided by expert domain knowledge.

\red Successful \nc model calibration is \red achieved \nc with data collected
from measurements of the \red phenomenon \nc under \red study\nc. For some
models, the collection of data can be so expensive that only a handful of
experiments is feasible. This is common in applications such as astrophysics
\citep{Vernon2014}, epidemiology \citep{Andrianakis2015}, and climate modeling
\citep{Salter2016}, to name a few. History matching
\citep{andrianakis2017efficient,Vernon2014} is a calibration technique that
iteratively discards regions of the input space through the use of an
implausibility measure. \red This way, \nc history matching overcomes the
limited availability of experimental data (measurements) and is able to identify
regions of input space that are likely to replicate the observed \red
phenomenon, \nc given different sources of uncertainty. Thus, the regions that
show high implausibility can be discarded from the analysis. Furthemore, history
matching is also able to determine whether there is no region of interest, that
is, no combination of input values that is likely to match the available data.
This can be used as evidence that further improvement in the simulator is
needed. This contrasts with typical Bayesian analysis of computer code output
(BACCO, \citep{Kennedy2001}) where positive posterior probability mass is always
assigned to regions that under the history matching framework would otherwise be
discarded.

Increasing complexity in mathematical models often translates in increasing
computational cost of the corresponding simulator. Thus, the ability to exploit
the simulator is limited by computational budget or time constraints. For
example, certain simulators used for climate models, nuclear reactor models, and
biological models \red need days to complete a single simulation run \nc
\citep{Smith2013}. This represents an additional layer of complexity, since the
ability to explore the input space is hindered by the high computational cost.
Common techniques such as Monte Carlo simulation and its variants are not well
suited in this context. In turn, fast but accurate emulators are needed to
overcome this limitation. In particular, Gaussian Process (GP) models have been
successfully used as Bayesian emulators in different scientific applications
such as machine learning \citep{Rasmussen2006}, spatial data analysis
\citep{Cressie1993}, genetics \citep{Kalaitzis2011}, and stochastic finite
element analysis \citep{DiazDelaO2011a}, to name a few. Since a GP provides a
full probabilistic characterisation of the unknown model output, its posterior
predictive mean provides a surrogate model, whilst its posterior predictive
variance measures the accuracy of the emulator.

The use of simulators and emulators introduces a wide range of uncertainties in
the modeling process. In history matching, these uncertainties are elicited and
incorporated in the variability of the predicted output
\citep{Craig1997,Vernon2010}. \red The implausibility measure is defined as a
function of the number of expected standard deviations between the observed data
and the corresponding emulator output \citep{Andrianakis2015}. In the
literature, it is common to use the mean of the emulator as an estimate within
the implausibility measure to guide the iterative selection of points in the
non-implausible domain \citep{Craig1997,Vernon2010,Wilkinson2014}. This is
achieved by computing the absolute value of the difference between the emulator
averaged prediction and the experimental data, standardised by different sources
of uncertainty. To the authors' knowledge, using the full probabilistic
characterisation of the emulator has not been explored for history matching
applications. \nc

In contrast, within the area of robust optimisation of black-box computer codes
\citep{Jones1998}, the probabilistic output of a GP is acknowledged and
incorporated in the exploration of the input space through the use of
appropriate acquisition functions \citep{Forrester2008a,Ranjan2008} . The
optimisation is performed as an iterative procedure that uses a GP model as an
emulator for the black box function \citep{Jones1998}. In these applications,
the acquisition functions incorporate the probabilistic information from the
emulator and is used in turn to guide the exploration of the input space. For
instance, the commonly used criterion of expected improvement is guaranteed to
find the optima of a function under certain regularity conditions
\citep{Vazquez2010}. As a consequence, expected improvement is often preferred
to deterministic estimates of improvement to guide the exploration of the input
space \citep{Forrester2008a}. This is because the expectation operator considers
the uncertainty modeled through the emulator, as opposed to using maximum
likelihood (MLE) or maximum a posteriori (MAP) estimate. Motivated by this
analogy, the history matching strategy developed in this paper uses a full
probabilistic characterisation of the output of the simulator to be able to
guide the reduction of the input space at every iteration.

As mentioned before, history matching is a sequential procedure that identifies
regions of non-implausible input configurations in order to refocus the
emulation of the simulator. Refocusing enables a better identification of
non-implausible points by using an improved emulator in the regions of interest.
This raises the question of how to choose new simulator runs to improve the
current emulator. In this paper, different functions to guide the selection of
points are tested. These functions are known in other research communities under
the name of active learning criteria \citep{eric2008active}, sampling criteria
\citep{Bect2012} and learning functions \citep{Lv2015}. In this work, the term
active learning is used in order to facilitate a link between the machine
learning and history matching communities. Moreover, the term active learning is
retained as these criteria guide the identification of regions where the limited
computational resources must be spent in light of the evidence of data and
observed response from the simulator.

\red This paper proposes the use of a full probabilistic characterisation of the
emulator within the implausibility measure to guide refocusing. \nc In
particular, three active learning criteria are generalised and presented as
choices to guide the selection of new training runs to iteratively improve the
emulator. Firstly, the expected contour improvement \citep{Ranjan2008} is used
as it was specially designed to refine an emulator for a given contour level.
Secondly, the expected risk \citep{Echard2013} used for reliability analysis is
modified here to adapt to contour estimation. Thirdly, the entropic profile
presented in \citep{Lv2015} is also modified to target a specific contour level
of an emulator.

The paper is organised as follows. In \cref{sec:history_matching} the
preliminaries for history matching are presented, with a brief overview of
GP emulators. In \cref{sec:nroy_id} the identification of
non-implausible regions is discussed within the context of the simulated
annealing sampling methods used in subset optimisation. This provides regions of
input parameter space that are likely to match the simulator output to observed
data. In \cref{sec:active_criteria} the proposed active learning criteria are
presented. In \cref{sec:experiments_hm} some illustrative examples are shown.
Concluding remarks are presented in \cref{sec:conclusion}.

\section{History matching} \label{sec:history_matching}

History matching is a calibration technique particularly useful in settings
where not only the model is computationally expensive, but the data-generating
process is expensive as well. The seminal papers \citep{Craig1996} and
\citep{Craig1997} introduced history matching within the framework of Bayes
linear statistics to analyse computationally expensive computer models.
\citet{Vernon2010} presented a thorough exposition of history matching in
large-scale high-dimensional applications such as the ones encountered in
cosmology. A more recent discussion of the history matching framework can be
found in \citep{Goldstein2016}.

In particular, this paper focuses on the application of history matching in
cases where the cost of generating new experimental data is so high that very
limited information from the physical process under study can be recorded. In
this setting, history matching aims to identify regions of the input parameter
space $\X$ of the simulator that are able to replicate the measured data, given
the structure of the model and the sources of uncertainty. This corresponds to a
relaxation of the search \red for a single optimal calibration point $\x^*$, \nc
that matches the simulator output to the physical process. This relaxation
consider regions of parameter space that are able to replicate the observed
process within a certain level of modeled uncertainty. \red The more stringent
alternative is the typical calibration setting where the objective is to find
the unique optimal configuration
\citep{Bayarri2007,tuo2016theoretical,Kennedy2001}. \nc\ In the typical Bayesian
alternative one would treat $\x^*$ as an unknown parameter, and the posterior
distribution for $\x^*$ would result from updating the prior specification given
some measurements from the physical process. However, it could be the case that
there is not enough information to believe that there is a unique choice for
$\x^*$. \red There might be doubts in one or several aspects of the specified
model. This could mean, for example, that the \red discrepancy between the model
and the phenomenon under study \nc -- its structure and independence from an
optimal $\x^*$ -- is not very well understood. These doubts make the
interpretation of $\x^*$ as the optimal calibration value meaningless. \nc\ In
this setting, history matching identifies collections of simulator evaluations
that are consistent with the measured data within the levels of uncertainty
associated with the problem.

The overall strategy of history matching is to use an emulator to explore the
input space in order to find regions on which the simulator gives acceptable
matches to the data. History matching then discards regions of input space (even
in cases where the simulator has a multi-dimensional output) in stages or {\it
waves}, as introduced in \citep{Vernon2010}. Thus, history matching sequentially
removes regions of parameter space using an implausibility measure. The input
regions that are considered non-implausible are sampled to refocus the emulator
for the next wave and, as a consequence, further reduce the non-implausible
region. The procedure of resampling, re-emulating, and reducing the
non-implausible space is done until a stopping condition is met. \red It should
be noted that the history matching process seeks to discard regions where poor
matches between simulator output and measurements. This can be done using only a
subset of outputs and observations, unlike other approaches which require
consideration of all simulator output coordinates. \nc

For the purpose of this paper, let $y$ denote the true physical process of
interest. Due to experimental error, $y$ cannot be observed directly. Let $z$
denote the noisy version of the process. That is, $z = y +
\epsilon_{\text{me}}$, where $\epsilon_{\text{me}}$ denotes an observational
noise with zero mean and finite variance. The limitation of not being able to
observe directly the data is what is commonly referred to as {\it observation
uncertainty} or {\it measurement error}. The quantity of interest, $y$, is
assumed to be the output of the simulator being calibrated. \red Let $f(\vec x)$
denote the simulator output using the input parameter $\vec x \in \X \subset
\R^d$. The simulator $f(\cdot)$ is assumed to be only a mathematical abstraction
of the true underlying process, which adds an additional layer of uncertainty in
the computer output. The inevitable mismatch between the computer model and the
process is called {\it model discrepancy} as in \citep{Kennedy2001}. Some
sources of model discrepancy are: reduced accuracy due to floating point
arithmetic, simplifying assumptions of the computational model, lack of
understanding of the underlying physics, among others. Let
$\epsilon_{\text{md}}$ denote the model discrepancy and assume $y = f(\vec x^*)
+ \epsilon_{\md}$, where $\x^*$ is the optimal calibration point. \red Note that
model discrepancy can be inferred simultaneously whilst performing calibration,
as it is done in the framework described by \citet{Kennedy2001}. However, due to
the confounding between $\x^*$ and the discrepancy term, it has proven useful to
infer first an appropriate calibrated emulator and then model the discrepancy
from the residuals. This is called modularized Bayesian inference of computer
code output \citep{Liu2009a,berger2019statistical}. In histroy matching,
however, this discrepancy is usually elicited from domain expert knowledge
\citep{Andrianakis2016}. \nc

As stated before, the computational complexity of the simulator inhibits the
ability to explore the configuration space. In typical industrial applications,
each run of the simulator could take as much as days or weeks to complete. As a
consequence, an additional layer of uncertainty is introduced. In the
literature, this is known as {\it code uncertainty}. An inexpensive
approximation for the simulator is used to cope with this limitation. In this
work, the emulator used for the simulator is a full Bayesian GP.
The use of a full Bayesian GP provides two advantages in history
matching applications. Firstly, uncertainty in the surrogate itself is partially
mitigated due to the marginalisation of the GP hyperparameters.
Secondly, as a by-product of GP emulators, code uncertainty can be
directly estimated due to the analytical expression for the output variability.

A GP is a nonparametric model used for Bayesian inference in
function spaces. A GP emulator considers the mapping from input space $\X$ to
output $y$ as a stochastic process indexed by $\X$. An intuitive way of thinking
of a GP is to view it as an infinite-dimensional extension of a multivariate
Gaussian distribution. Just as its finite dimensional counterpart, it is
completely determined by its mean function $m(\cdot)$ and covariance kernel
$k(\cdot, \cdot)$, which specify its first two moments. The mean function
$m(\cdot)$ embodies the understanding of any global trend exhibited by the true
process. The covariance kernel, on the other hand, incorporates prior knowledge
of any assumptions on how similar input configurations $\x,\x' \in \X$ produce
correlated outputs. An additional property of GP emulators is that the model
produces its own measurement of variability, commonly used as predicted error.
\red GP emulators have become standard tools to quantify uncertainty in
expensive computer models. For more detail on their theoretical underpinnings
and implementation, the interested reader is referred to
\citep{Rasmussen2006,Oakley2004}. \nc

Let $\sigma(\x)^2 = k(\x,\x)$ denote the variability of predicted code output at
configuration $\x$. \red In order to perform Bayesian inference for the GP, the
mean and covariance functions have to be chosen beforehand from a family of
possible choices \citep{Rasmussen2006}. It is common to specify a zero mean
process with a squared exponential kernel, which for simplicity is done in this
paper. \red This choice corresponds to a GP that can be interpreted as a radial
basis expansion on the locations of the training data. This covariance kernel
assumes the emulated function to be infinitely differentiable. \nc Note that
elicitation of an appropriate mean function and covariance kernel can be done in
such a way that it incorporates domain expert knowledge on expected behaviour of
the simulator \citep{Oakley2002a,Vernon2010}. The mean and covariance functions
are further parameterised by its respective vectors of hyperparameters. These
hyperparameters are often selected by an empirical Bayes approach. However in
this paper, a fully Bayesian procedure is used in order to acknowledge the
limited number of simulator runs available. More details are given below.

Let $\mathcal{D} = \{(\x_i, y_i)\}_{i = 1}^N$ be the collection of input--output
pairs used to train the GP emulator. This collection of points is usually chosen
so that the input configuration space is explored uniformly in every dimension.
A Latin Hypercube Sampling (LHS) scheme is usually chosen to this end. Bayesian
inference incorporates prior knowledge on the hyperparameters of the
GP, if available, and allows one to compute the posterior distribution given the
observed training data. Given the training runs and a Gaussian measurement model
for $\epsilon_\me$, it can be shown that the posterior prediction on an unseen
input configuration $\x^*$ follows a Gaussian distribution with posterior mean
and covariance functions given by
\begin{align}
m(\x^*) &= \sum_{i=1}^N w_i \,\, m_i(\x^*), \label{eq:montecarlo_prediction} \textbf{}\\
\text{cov}(\x^*,\x') &= \sum_{i=1}^N w_i \,\, \left[ (m_i(\vec{x}^*) - m(\vec{x}^*))
(m_i(\x') - m(\x')) + \text{cov}_i(\x^*,\x') \right], \label{eq:montecarlo_variance}
\end{align}
where $m_i(\x^*)$ and $\text{cov}_i(\x^*,\x')$ denote the posterior mean and
covariance functions of the GP emulator with hyperparameter vector $\boldsymbol
\theta_i$. The sum denotes a Monte Carlo approximation to the integral with
respect to the posterior distribution $p(\boldsymbol\theta | \mathcal{D})$ with
corresponding weights $w_i$. \red Note that $\text{cov}(\x^*,\x^*) =
\sigma^2(x^*)$ denotes the predicted variance of the GP prediction marginalised
by samples of the posterior distribution of the hyperparameters.\nc\ For a more
detailed explanation, the interested reader is referred to \citep{Garbuno2016}
and \citep{Rasmussen2006}.

Let $I(\x)$ denote the implausibility measure of the input configuration $\x$
given the observed datum $z$. This implausibility is defined as \red
\begin{align}
I(\x) = \frac{\Big|z - m(x) \Big|}{\sqrt{\sigma^2(\x) + \sigma^2_{\text{md}} + \sigma^2_{\text{me}}}} \label{eq:implausibility},
\end{align}
where $m(x)$ and $\sigma^2(x)$ denote, respectively, the posterior mean and
posterior variance of the GP emulator as defined above.\nc\ It is important to
note that, in certain applications, the simulator output is known to be
stochastic and an additional term is added in the denominator to account for
{\it ensemble variability} \citep{Andrianakis2015}. In this work such
variability is not needed. Furthermore, note that the distance between the data
threshold $z$ and the surrogate output is standardised by the sum of the modeled
uncertainties. The implausibility function helps identify which configuration
points are far from the target, as measured by a number of standard deviations.
In the literature, Pukelsheim's three sigma rule \citep{Pukelsheim1994} is a
common choice to characterise the number of standard deviations in this setting.
The rule states that if $X$ is a \red continuous \nc random variable with mean
$m$ and variance $\sigma^2$ which follows a uni-modal distribution, then the
probability for $X$ falling away from its mean by more than 3 standard
deviations is at most 5\%. That is,
\begin{align}
P\{|X - m| > 3\, \sigma\} < 0.05
\end{align}

Following the above criteria, the region of input space where the emulator
should refocus is then defined as
\begin{align}
\X_{\text{NROY}} = \{ \x \in \X \,:\, I(\x) \leq k \}\label{eq:nroy},
\end{align}
\red where $k = 3$ following the above considerations, and the subscript stands
for {\it Not-ruled-out-yet} (NROY) \citep{Vernon2014}.\nc

\subsection{History matching with probabilistic emulators}

The previous description of history matching stems from the construction of
probability by using mathematical expectation as a primitive. This approach is
known as Bayes linear statistics, since linearity of expectations is a key
aspect in the development of tools under the theory. For a more thorough
discussion on Bayesian linear methods refer to \citep{Goldstein2007}. The
construction of the implausibility measure under Bayes linear can be seen as a
function that uses a {\it pointwise} estimate from the emulator. Nonetheless,
the GP emulator provides a probabilistic generator of surrogate models, which
can be exploited in a full probabilistic formulation. \red As discussed above, a
full probabilistic approach is proposed in this paper, whereby both the GP
probabilistic output and the posterior distribution of the hyperparameters are
considered for the emulator. \nc

\red In the proposed characterisation, a full probabilistic treatment of the
emulator is used within the implausibility function. This is achieved by
incorporating the probabilistic distribution of the GP emulator output. Let
the emulated implausibility be defined as
\begin{align}
\Igp(\x) = \frac{|z - \fgp(\x)|}{\sqrt{\sigma^2(x) + \sigma^2_{\md} +
\sigma^2_{\me}}}, \label{eq:imp_prob}
\end{align}
where $\fgp(\x) \sim \N(m(\x),\sigma^2(\x))$ is the GP emulator for the
simulator output. The dependence on the training runs $\mathcal{D}$ is
omitted to avoid cluttered notation. The NROY space is thus characterised by
probabilistic statements of the form
\begin{align}
P\{ \Igp(\x) \leq 3\}, \label{eq:nroy_charact}
\end{align}
which is analogous to the probabilistic statements of
\citep{Holden2015} and \citep{Williamson2013}. \nc

\begin{figure}[!htp] \centering \subfigure[Implausibility measure based on a deterministic
emulator.]{\includegraphics[width=.45\linewidth]{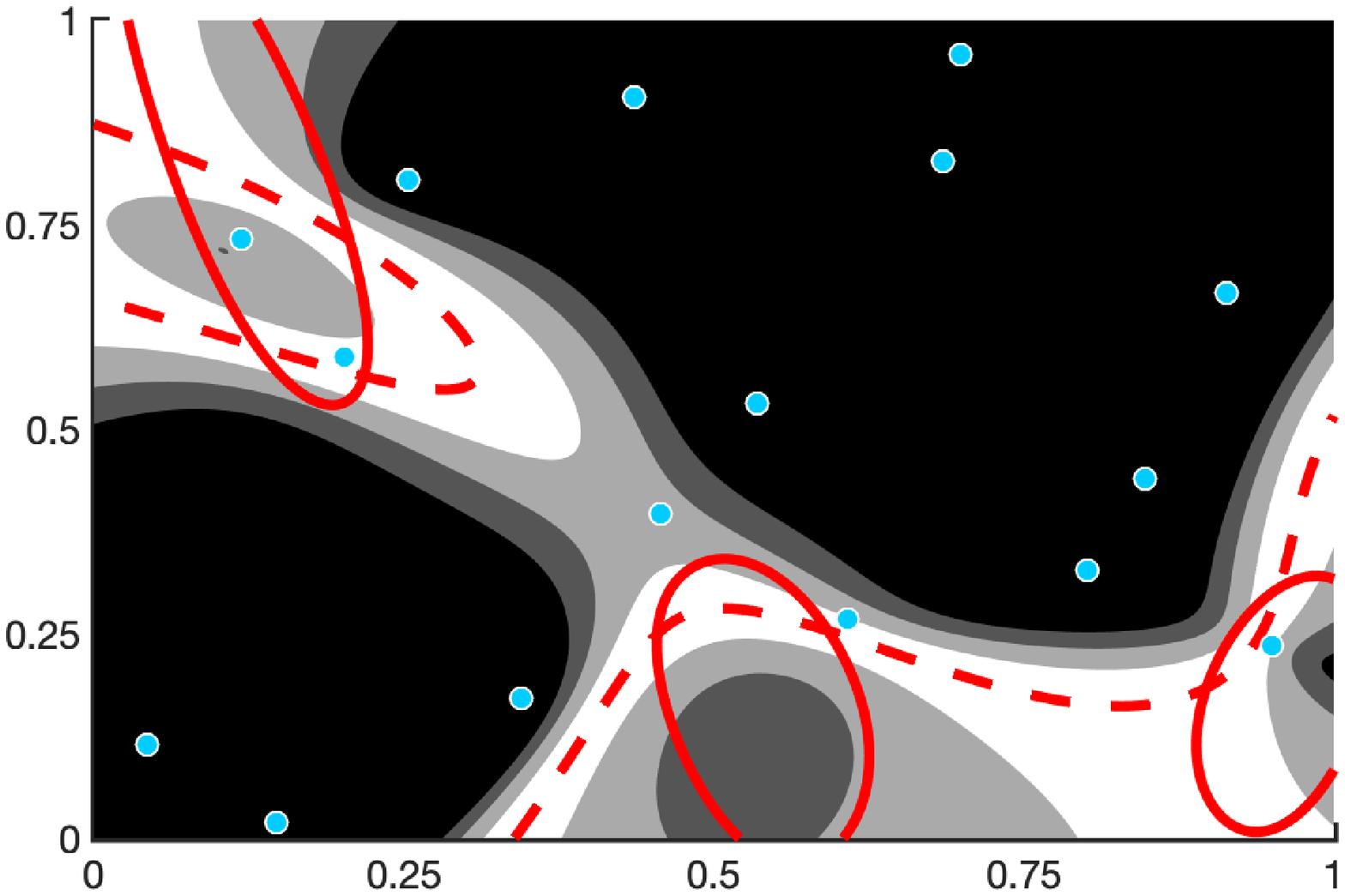}
\label{subfig:imp_pnt}} \qquad \subfigure[Probability of implausibility based on stochastic
emulator.]{\includegraphics[width=.45\linewidth]{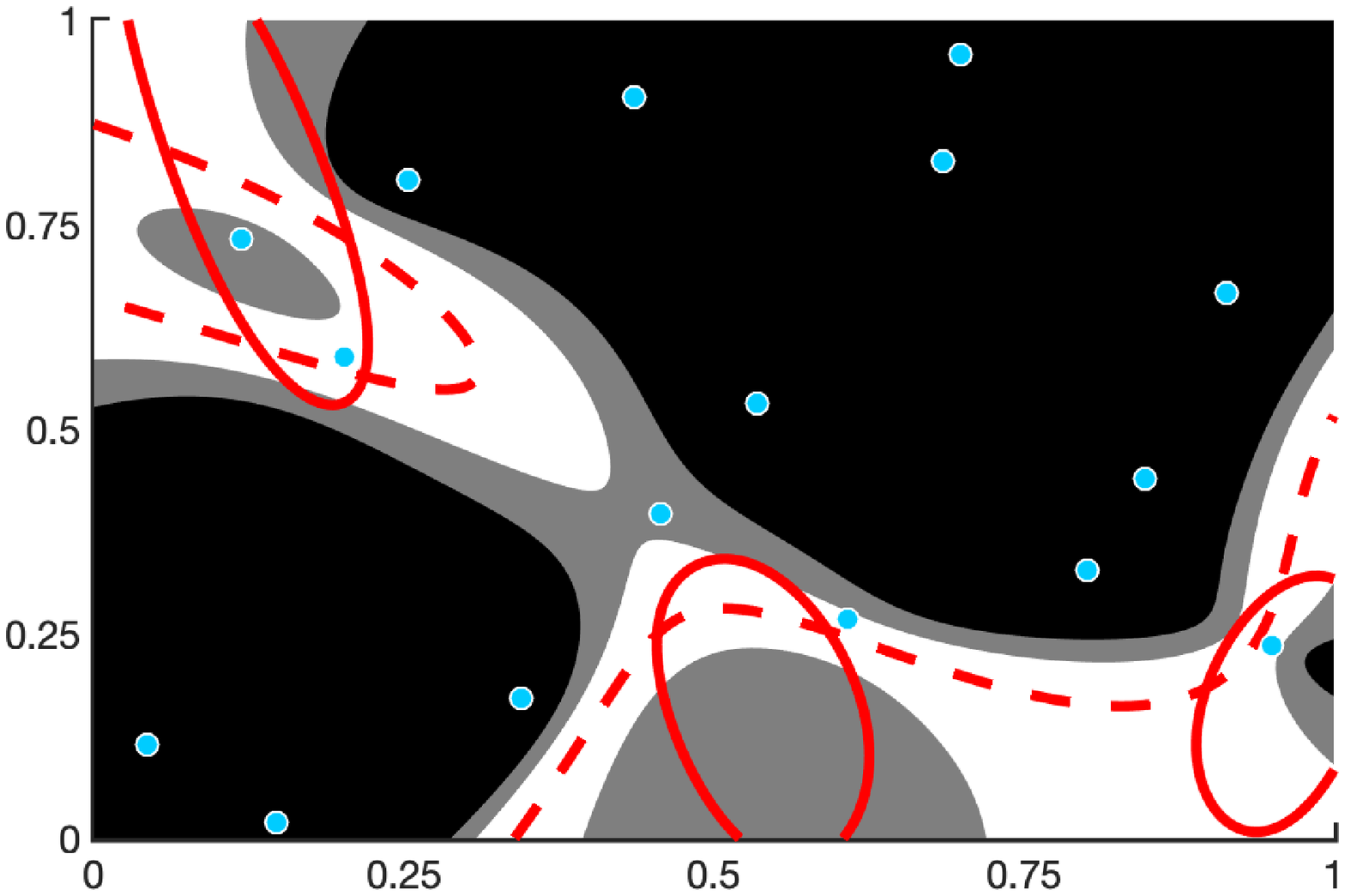}
\label{subfig:imp_emu}}

\caption{\red Illustration of the contour levels for the implausibility function
around a target level (solid red line) of the Branin funcion. The GP emulator,
trained with the samples shown as dots, provides the dashed lines as its
prediction. The left panel shows the implausibility measure using a
deterministic emulator, the GP mean. The right panel shows the probability of
implausibility derived from the stochastic emulator.\nc}

\label{fig:implausibility_compare}
\end{figure}

The difference between using \eqref{eq:implausibility} and \eqref{eq:imp_prob}
is depicted in \cref{fig:implausibility_compare}. The underlying \red model \nc
is the modified Branin function \citep{Forrester2008a}, for which a GP emulator
is fitted using the blue dots as training points. Dashed lines correspond to the
emulator's response at the target level $z = 10$, whilst solid lines correspond
to the target level of the Branin function. In both subfigures, the dark-shaded
regions correspond to higher values of implausibility, whereas light-shaded
regions account for lower values. \red The contour levels correspond to the
number of standard deviations away from the target level $z$. Pitch-black
regions in \cref{subfig:imp_pnt} indicate values of the implausibility function
of $3$ and decrease one unit at a time as the colour becomes white. In
\cref{subfig:imp_emu}, the levels are chosen logarithmically so that the
pitch-black level corresponds to the region of more than $99\%$ probability.
This is dictated by the emulator predictive distribution. The rest of the
contour levels in \cref{subfig:imp_emu} decrease to levels $90\%$ and $67\%$.

The stochastic nature of the emulator motivates the direct description of
implausibility in terms of probabilities. This formulation naturally
incorporates the variability from the emulator, which is highly convenient for
computationally expensive simulators. In this kind of setting, only a small
amount of training data is available, and thus a fully Bayesian approach to
emulation might be desired. The dichotomy of selecting new simulator runs close
to the emulator response contour level or where there is high uncertainty, is
known as the exploration-exploitation trade-off in computer experimental design
\citep{Forrester2008a}. \nc\ Exploration is desirable as the use of an emulator
might induce bias if followed too blindly in the first steps of the procedure.

\section{NROY space identification} \label{sec:nroy_id}

History matching relies on the correct identification of the region of input
space $\X$ where the simulator is likely to replicate the observed data. At
every iteration, the NROY space becomes orders of magnitude smaller than the
original space, and can exhibit a complex or disconnected topology. As a result,
naive rejection-based sampling can quickly become \red very inefficient. \nc To
address this deficiency, different alternatives have been presented in the
literature. \citet{Williamson2013a} proposed an algorithm in the spirit of
simulated annealing. It is an implausibility driven sampling scheme which needs
to define an appropriate threshold ladder. \citet{Yeh2016} use clustering to
identify possibly disconnected regions. \citet{Andrianakis2015} proposed to use
Gaussian random variables centred at the mean from the NROY points of wave $t$
to generate points for wave $t+1$. For this, the covariance matrix is chosen so
that much of the input space can be covered, ideally accepting 20\% of the
proposed samples. \red Other recent approaches have been proposed by
\citep{andrianakis2017efficient}, who use a slice sampling approach to sample
from within the NROY space; \citep{drovandi2017new}, who solve the sampling
problem with Sequential Monte Carlo (SMC) and global information such as the
empirical covariance; and \citep{Gong2016}, who propose the use of a rare-event
sampling strategy usually employed in a reliability analysis context, namely
subset simulation.\nc

At present, the correct identification of the NROY space in a full probabilistic
setting has not been fully explored. \red To address this limitation, this paper
proposes the use of a sampling scheme that is able to generate approximate
independent samples from the target region, even when the NROY space exhibits
challenging features such as being disconnected.\nc\ Inspired by sequential
Monte Carlo, simulated annealing and subset stochastic optimisation,
\citep{Beck} developed an algorithm (AIMS) that draws approximate independent
samples on a set of interest. \red A variation of AIMS was then published in
\citep{Zuev2013}. The algorithm (AIMS-OPT) was shown to achieve excellent
results in complex stochastic optimisation settings, \eg when the maximum can be
achieved in a ridge on the input space. This later motivated the development of
the algorithm called \tas \citep{Garbuno2016}, where a modification was proposed
to improve efficiency through slice sampling, as well as by exploiting
parallelisation. \tas is the sampling scheme used throughout this paper. Full
details of the algorithm can be found in \citep{Garbuno2016}, and the interested
reader can access a full implementation in the repository
\url{https://github.com/agarbuno/ta2s2_codes}.\nc

In this work, the focus is on regions in which the probability of
non-implausibility in input space $\X$ is maximal. That is, we aim for the
maximisers of $P\{ \Igp(\x) \leq 3\}$, which are hopefully close to the upper
bound of 1. \red This objective corresponds to the three sigma rule mentioned in
Section 2, which states that, for continuous unimodal distributions, $95\%$ of
probability is achieved within three standard deviations from the mean. \nc\ By
means of the \tas algorithm in \citep{Garbuno2016} a nested sequence of sample
sets $U_m \subset \ldots \subset U_0$ is obtained such that \red
\begin{align}
U_j = \left\lbrace \x_i^{(j)}: \x_i^{(j)} \sim p_j(\x), \, i = 1, \ldots, N\right\rbrace,
\end{align}
\nc where $p_j$ denotes an intermediate density that converges to a uniform density
in the set of optimisers, and $N$ is the number of samples extracted at every
level of the annealing schedule \citep{Zuev2013, Garbuno2016}.

One of the \red key \nc advantages of using \tas in history matching, \red as
opposed to other sampling schemes, \nc is that if the resulting set $U_m$ is
highly concentrated at one probability level, the previous level of samples can
be used instead for exploration. This would be convenient if more samples from
lower probability responses are needed. \red Consider the case where most of the
samples drawn within the set $U_m$ provide a very highly concentrated collection
of values $P\{ \Igp(\x) \leq 3\}$. This could signal the presence of points very
close to a neighborhood of a previously tested simulator run. Thus, if desired,
additional subsets $U_j$ for $j < m$ can be explored. This of course, can be
problem-dependent and might need careful additional considerations. \nc\

\begin{figure}[t]
\centering
\includegraphics[width=.65\linewidth]{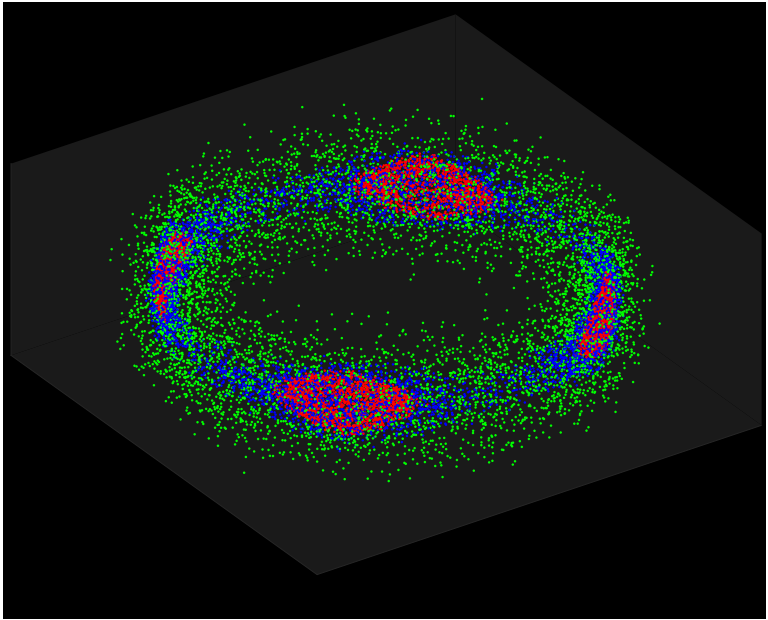}

\caption[Torus implausibility exercise for multimodal sampling]{Samples
generated at each wave for the torus implausibility function of
\citep{Williamson2013a}. At the final level, the adaptive sampler correctly
identifies the four regions of the zero-th contour level (in red). It should be
noted that exploration of the space could be done with samples from previous
levels if required.}

\label{fig:torus}
\end{figure}

In Figure \ref{fig:torus} an application of the \tas algorithm is shown to
adaptively identify the NROY space for a torus example presented in
\citep{Williamson2013a}. The function is defined over the 3-dimensional cube
$[-20,40]^3$ and its expression is as follows. Let $\x = (x_1, x_2, x_3)^\top$,
and define the 2-dimensional projection as
\begin{align}
\boldsymbol u = \left[
\begin{array}{c}
(x_1-2)^2 - 3 \\ (x_2-2)^2-3
\end{array}
\right], & & \Sigma = \frac{1}{2^{12}}\left(
\begin{array}{cc}
1 & -0.97\\-0.97 & 1
\end{array}
\right).
\end{align}
The implausibility function for this numerical exercise is defined as
\begin{align}
I(\x) = \frac{1}{10} \left( \sqrt{\boldsymbol u^\top \Sigma^{-1} \boldsymbol u} + \frac{x_3^2}{0.04^2}\right).
\end{align}
which induces the 4 the modes in a
torus as shown in \cref{fig:torus}. For this numerical exercise, the only
assumed source of uncertainty is the measurement error as the exact model is
being used in the search, \ie no emulator was necessary.

\section{Sequential non-implausible design} \label{sec:active_criteria}

Having successfully identified the NROY space, the question of how to query from
such region to refocus the emulator remains a challenging problem. The seminal
papers \citep{Craig1996} and \citep{Craig1997} performed a sequential design for
this purpose. The common choice in the literature is to select points greedily
from the NROY space to build a new emulator completely focused on that region
\citep{Salter2016}. This implies that a new emulator is built based on the
identified region at every wave. In this work, a different (albeit conservative
approach) is followed. Since the simulator is assumed to be computationally very
expensive, it might seem unrealistic to expect that the computational budget is
kept the same at every wave. Moreover, discarding points might represent a waste
of resources and information. In turn, the points must be chosen carefully at
each wave to later add them to the set of training runs and build a new GP
emulator.

The idea is that the most general information would likely be extracted at the
very first iterations while greater accuracy will be pursued at later stages of
the history matching procedure. For example, in the first waves a good
characterisation of the global trend can potentially be identified. It is
therefore appropriate to guide the choice of training points by following
suitable learning criteria. This is done in Bayesian optimisation or in
reliability analysis problems \citep{Bect2012,Kuczera2009,Villemonteix2009}. For
comparison, three active learning criteria are studied.

The first criterion is the expected
contour improvement by \citep{Ranjan2008}. The improvement function is defined
as \red
\begin{align}
\mathcal{I}_z(\x) = \epsilon^2(\x) - \min\left\lbrace (\fgp(\x) - z)^2, \epsilon^2(\x)\right\rbrace,
\end{align}
where $\fgp(\x)$ is the GP emulator at configuration $\x$, $z$ the targeted
contour level, and $\epsilon(\x) = k \, v(\x)$ the number of predicted standard
deviations derived from the uncertainty model, $v(\x) = \sqrt{\sigma^2(\x) +
\sigma^2_{\md} + \sigma^2_{\me}}$. Note that the full uncertainty model is
employed in order to be consistent with the preceeding notation. In practice,
since both measurement error and model discrepancy are assumed to be constant,
this assumption does not affect the history matching pipeline considered here.
In the context of simultaneous model calibration and discrepancy learning as in
\citep{Bayarri2007,Liu2009a} more careful considerations should be taken. For
instance, the above points can be addressed by using only the predicted
variability of the GP, since the approach developed here aims at improving the
emulator. \nc In this setting, $k$ can be chosen as 3, following the three sigma
rule. The expected value of the contour improvement is used, given that the
emulator is random in nature. The expected contour improvement (ECI) can be
computed as
\begin{align}
\mathbb{E}[\mathcal{I}_z(\x)] = &
\left[ \epsilon^2(\x) - (m(\x) - z)^2 - \sigma^2(\x) \right] \left[\Phi(z_2) - \Phi(z_1)\right]\nonumber\\
& + \sigma^2(\x) \, \left[z_2 \, \phi(z_2) - z_1\,\phi(z_1)\right] \label{eq:eci} \\
& + 2 [m(\x) - z] \, \sigma(\x) \left[\phi(z_2) - \phi(z_1)\right] \nonumber
\end{align}
where $z_1 = (z - m(\x)-\epsilon(\x))/\sigma(\x)$, $z_2 = (z -
m(\x)+\epsilon(\x))/\sigma(\x)$, and $\Phi(\cdot)$ and $\phi(\cdot)$ are the
standard Gaussian cumulative and density functions respectively.

The second criterion to be considered is the expected risk by
\citep{Echard2013}, which was originally designed for reliability analysis. This
\red aims at \nc learning the set $\{\x : g(\x) > 0\}$, with $g(\x)$ the
performance or limit-state function of a configuration $\x$. The critical level
$g(\x) = 0$ is referred to as a {\em transition} level, as its correct emulation
classifies a given configuration in terms of the system's performance. In this
paper, the problem is explicitly stated in terms of the contour level $z$ which
corresponds to the observed data in the experimental setting. This means that
the risk function is defined as
\begin{align}
R_z(\x) = \left\lbrace \begin{array}{ll}
(\fgp(\x) - z)_+ & \text{if } m(\x) \leq z \\
(z - \fgp(\x))_+ & \text{if } m(\x) > z
\end{array} \right.,
\end{align}
where $(\cdot)_+$ denotes the non-negative part of the argument, and $m(\x)$
denotes the expected value of the GP emulator at configuration $\x$. The
expected risk is used as a learning criterion due to the random nature of the
output of the emulator. The derivation is a straightforward solution of one
dimensional Gaussian integration, which for completeness is included in
\cref{apx:risk_solution}. The analytical expression can be written in compact
form as
\begin{align}
\mathbb{E}[R_z(\x)] = \sigma(\x) \left[ -\text{sign}(\bar{z}) \, \bar{z} \,
\Phi\left(-\text{sign}(\bar{z}\right) \, \bar{z}) + \phi(\bar{z})\right],
\label{eq:risk}
\end{align}
where $\bar{z} = (z - m(\x)) / \sigma(\x)$ denotes the standardised
contour level, $\text{sign}(\cdot)$ the sign function, and the pair
$\Phi(\cdot)$ and $\phi(\cdot)$ are the cumulative and density functions
used as before.

The third learning criterion to be compared is a variation of the
entropic profile presented by \citep{Lv2015}. Originally formulated in the
reliability analysis literature, it was designed to measure the entropy of a
random variable in a neighbourhood of two standard deviations from the origin.
In this paper, the concept has been extended. Again, an explicit solution is
presented for a contour level $z$ observed in the experimental data. The
entropic profile is defined as
\begin{align}
H_z(\x) = \Bigg|\int_{z-k\,\sigma(\x)}^{z+k\,\sigma(\x)} - \ln \pi(\fgp)\,\pi(\fgp)\,\rd \fgp \Bigg|.
\end{align}
As shown in \cref{apx:entropy_solution}, the entropic profile can be written
compactly as
\begin{align}
H_z(\x) = \Big| \left[ \ln\left(\sqrt{2\pi}\,\sigma(\x)\right) + 0.5\right] \,
\left[\Phi(z_2) - \Phi(z_1)\right] - 0.5 \, \left[ z_2
\phi(z_2) - z_1\phi(z_1)\right]\Big|,
\label{eq:entropic_profile}
\end{align}
where, as before, $z_1$ and $z_2$ denote the standardised contour levels.

All the above learning criteria rank the samples from the identified NROY space.
It is important to note that this type of criteria take into account a
one-step-look-ahead pointwise strategy. Other options include A-optimal designs,
which incorporate area impacts to the improvement of the emulator's response
surface. See \citep{Chaloner1995} for a thorough discussion of optimal design of
experiments. In particular, following dynamic programming strategies, one can
define a learning criteria with a known number of sequential decisions. As a
consequence, this type of selection of points choose a batch of candidate runs.
This is known as finite-horizon dynamic programming \citep{Bertsekas2005}, and
is subject of future study which falls outside the scope of this paper.

Given the proposed learning criteria, a natural question is how to choose the
points in the resulting ranking. Since nearby sample points are likely to be
similarly ranked, it would be naive and a waste of computational resources to
query the simulator on just the top-ranked NROY samples. Alternatively, to
choose uniformly from the sampled points would ignore any ranking at all. An
adequate leverage between these extremes is achieved by the following procedure.

Firstly, after a learning criterion has been chosen to rank the samples, a cut
off point needs to be selected to specify a subset of good candidates. This cut
off is a threshold to guarantee that some percentage of the maximum attainable
gain can be held. \red In this work, this threshold is set to be $50\%$. The
rational for this choice is the following. For a very high percentage cut off,
the strategy would likely concentrate in narrow neighbourhoods around the
highest-ranked sample. In contrast, a very low percentage would not acknowledge
the ranking at all. To strike a balance, a conservative approach is to only
include those samples able to attain at least a $50\%$ best score. The optimal
value for this threshold could be problem-dependent, and a more detailed
numerical study is the subject of future research. Once this threshold is set,
the first point chosen is the one ranked the highest by the active learning
criterion. By construction, the rest of the points above the 50\% threshold will
have a lower ranking, but will still be desirable samples. In practice, this
means that the learning criteria will guide the first points in the rank, whilst
the rest should be sampled following a space filling design in order to train
the emulator.

Secondly, a maximin design is proposed to choose the next batch of training
points from the collection of desirable training runs, that is, points above the
cut off value. The seed of this design is selected to be the top ranking point
from the NROY samples, \emph{i.e.} the point that has the highest expected
gain. \nc It is important to note that this type of sampling usually starts with
the mean of a cloud of points \citep{Kuczera2009}. Since the NROY space has
potentially a complicated, possibly disconnected topology, the average point
might lie outside the region of interest and choosing a point from outside the
NROY space could potentially be a poor selection.

\red Concretely, the proposed strategy proceeds as follows. After choosing the
starting point $\x_0$, construct the new set of training points as
$\mathcal{D}^* = \{\x_0\} \cup \mathcal{D}$. The next point selected for the
maximin design is the furthest sample available within the NROY space
weighted by the cutoff point $\alpha$, denoted by $\X_{\text{NROY}}(\alpha)$, so that
\begin{align}
  \x_1 = \underset{\x \in \X_{\text{NROY}}(\alpha) \setminus \mathcal{D}^*}{\arg\max} \| \x - \x_0\|,
\end{align}
is chosen as a successor. The new training data is updated accordingly,
$\mathcal{D}^* \leftarrow \{\x_1\}\cup \mathcal{D}^* $. The next step computes
distances of the sampled points in $\X_{\text{NROY}}(\alpha)$ to the set
$\mathcal{D}^*$ and retains the candidate farthest apart from $\mathcal{D}^*$.
This is done iteratively by computing
\begin{align} \label{eq:maximin}
  \x_k = \underset{\x \in \X_{\text{NROY}}(\alpha) \setminus \mathcal{D}^*}{\arg\max} d(\x,\mathcal{D}^*),
\end{align}
where
\begin{align}
  d(\x,\mathcal{D}^*) = \min_{\x' \in \mathcal{D}^*} \| \x - \x'\|,
\end{align}
denotes the distance of $\x$ to the current set of training data
$\mathcal{D}^*$. After every point is found by \eqref{eq:maximin},
the training dataset is updated accordingly $\mathcal{D}^* \leftarrow \{\x_k\}
\cup \mathcal{D}^*$. This procedure is repeated sequentially until $N$ new
samples to train a refocused emulator are gathered. As mentioned above, the
first point will follow the specific learning criterion chosen. The rest of
selected points will followed a space-filling design. \nc

In summary, the use of the maximin selection allwos one to (i) retain the best
possible point; (ii) collect samples from a space-filling design in NROY space;
and (iii) restrict the choice of new points following the active learning
criteria.

\section{Numerical experiments} \label{sec:experiments_hm}

In this section, the performance of the proposed \red history matching approach
is applied to \nc a 2D example, a 3D case study of a fault model, and a battery
of multidimensional tests. The 2D example serves as an illustration of the
proposed approach. In particular, the maximin design to choose from the ranked
samples. The 3D example illustrates a multi-output use of history matching.
Finally, the testbed of random functions provides a setting where the approach
is tried in different dimensional settings.

In history matching applications, a common stopping rule to terminate the
procedure is to compare the maximum predicted error of the emulator in the NROY
samples to the estimated variance attributed to measurement and model
discrepancy ($\sigma_{\me}^2$ and $\sigma_{\md}^2$). The motivation is that
further improvement of the surrogate would not be able to reduce the elicited
deviations from the simulator. The GP, in this setting, is ideal since it
provides an estimation of predicted error as a by-product of its construction.
As an alternative to this criterion, this work explores the use of a scoring
rule, the Continuously Ranked Probability score (CRPS) \citep{Grimit2006}. It
has the properties of being a proper scoring rule to report probabilistic
inferences. As stated before, the GP emulator is able to provide full
probabilistic statements like predicted values and dispersion estimates around
such predictions. \red The benefit of using the CRPS over local scoring rules
like the negative logarithm of predictive density (NLPD) relies on the fact that
localised rules risk penalising heavily over-confident predictions whilst
treating under-confident and far-off predictions more leniently. \nc In
contrast, the CRPS aims for better placement of probability mass near target
values, although not exactly placed at the target. The interested reader is
refered to \citep{Grimit2006} and \citep{Kohonen2006} . In particular, it is
known that the full Bayesian treatment in GPs is preferred for better error
estimation in uncertainty analysis \citep{Kennedy2001} and thus CRPS provides an
appropriate scoring rule for GP emulation. \red This exploits the fact that
under the Monte Carlo approximation, the predicted value and variance of the GP
emulator is a mixture of Gaussians, as seen in \eqref{eq:montecarlo_prediction}
and \eqref{eq:montecarlo_variance}. The CRPS evaluated at $\x$ corresponds to
\begin{align}
\text{CRPS}\left( \sum_{k = 1}^N \omega_k \, \mathcal{N} \left( m_k(\x), \sigma^2_k(\x) \right)\, , \, z \right) = &
\sum_{k = 1}^N \omega_k \, A \left(z - m_k(\x), \sigma^2_k(\x) \right) \nonumber \\
& \quad - \frac{1}{2} \sum_{k = 1}^N \sum_{l = 1}^N \omega_k \omega_l \, A\left(m_k(\x) - m_l(\x), \sigma^2_k(\x) + \sigma^2_l(\x)\right),
\end{align}
where $\omega_k$ denotes the weight of $k$-th Gaussian component of the mixture,
and $m_k(\x)$ and $\sigma^2_k(\x)$ are the corresponding mean and variance of the GP
emulator evaluated at the index $\x$. The function $A(\cdot, \cdot)$ is defined
as
\begin{align}
A(m,\sigma^2 ) = 2 \sigma \phi\left( \frac{m}{\sigma}\right) +m \left( 2 \, \Phi \left( \frac{m}{\sigma}\right) -1 \right),
\end{align}
where $\phi(\cdot)$ and $\Phi(\cdot)$ denote the density and cumulative
functions of a standard Gaussian random variable. It should be noted that the
CRPS is measured in the same units as the output of the simulator and can be
evaluated at every location of interest. In this work, we measure
the GP predictive capabilities over the NROY space through the samples generated
from the strategy described in \cref{sec:nroy_id}. The interested reader is
referred to \citep{Grimit2006} for more properties of the CRPS.\nc\

\subsection{Franke's function}

\begin{figure}[!ht]
\centering
\subfigure[Contour levels and training runs]{
\fboxsep=0mm
\fbox{\includegraphics[keepaspectratio = true, width = .3\linewidth, trim = 75 0 75 1, clip]{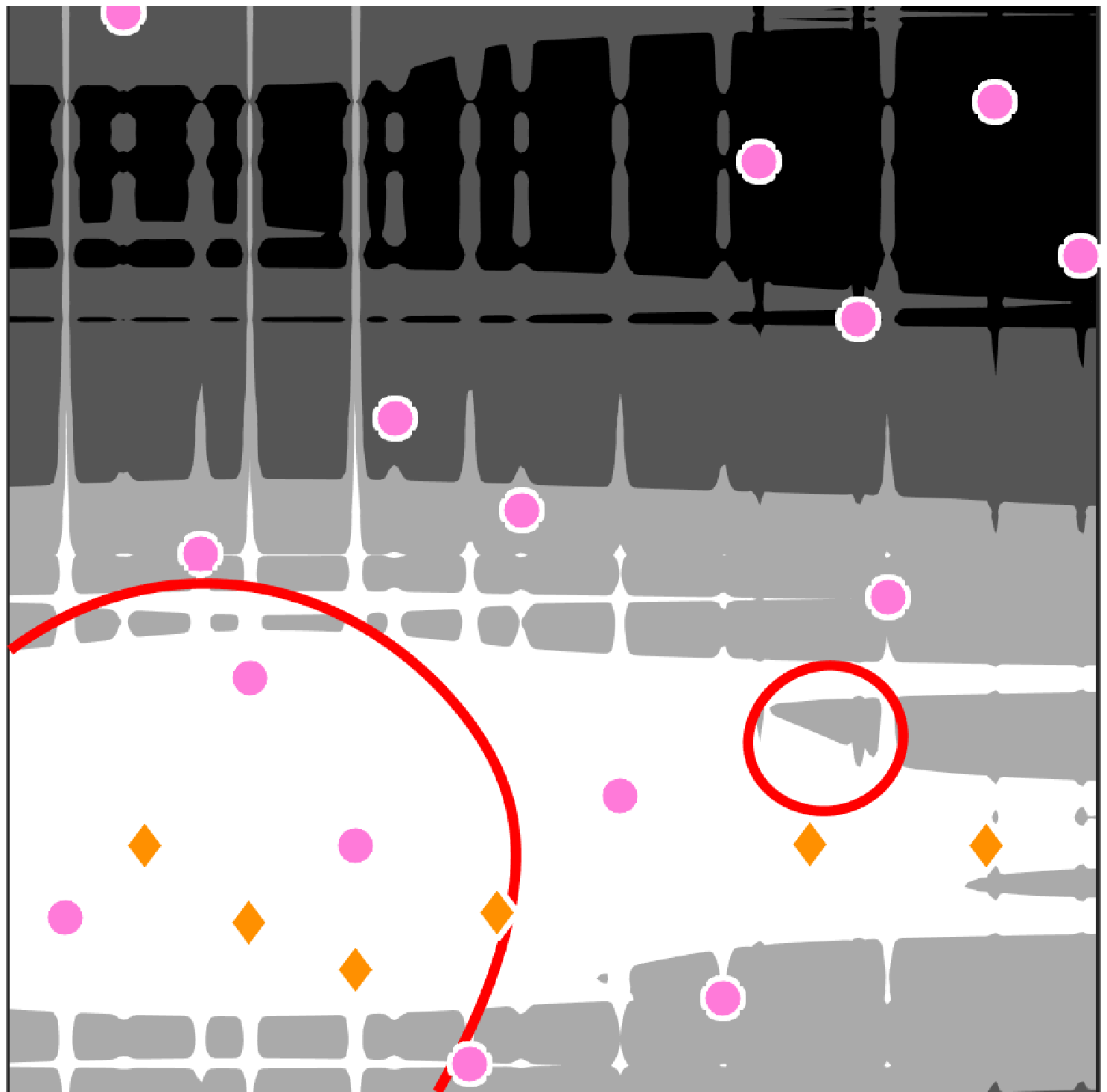}}\qquad
\fbox{\includegraphics[keepaspectratio = true, width = .3\linewidth, trim = 75 0 75 1, clip]{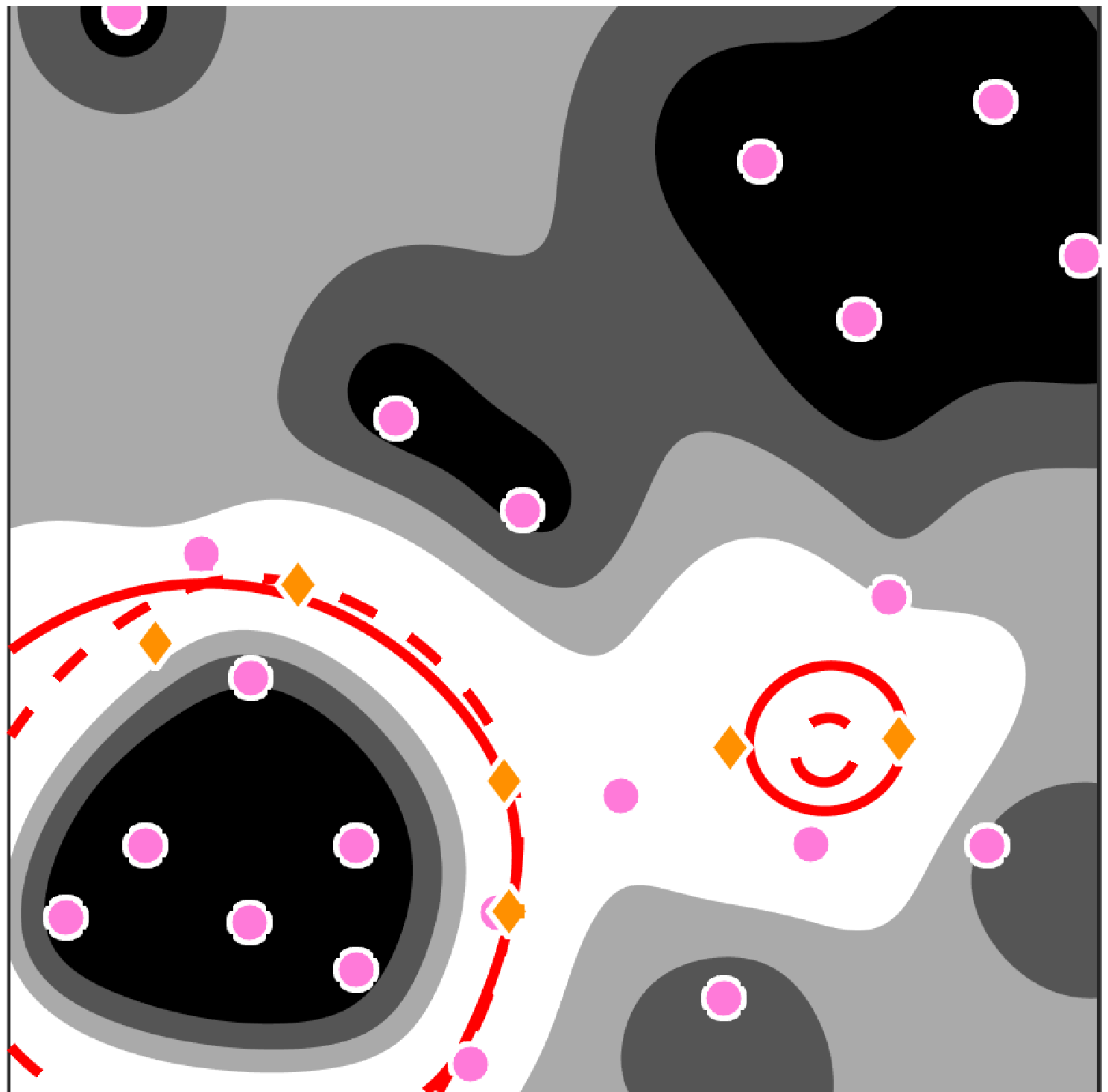}}\qquad
\fbox{\includegraphics[keepaspectratio = true, width = .3\linewidth, trim = 75 0 75 1, clip]{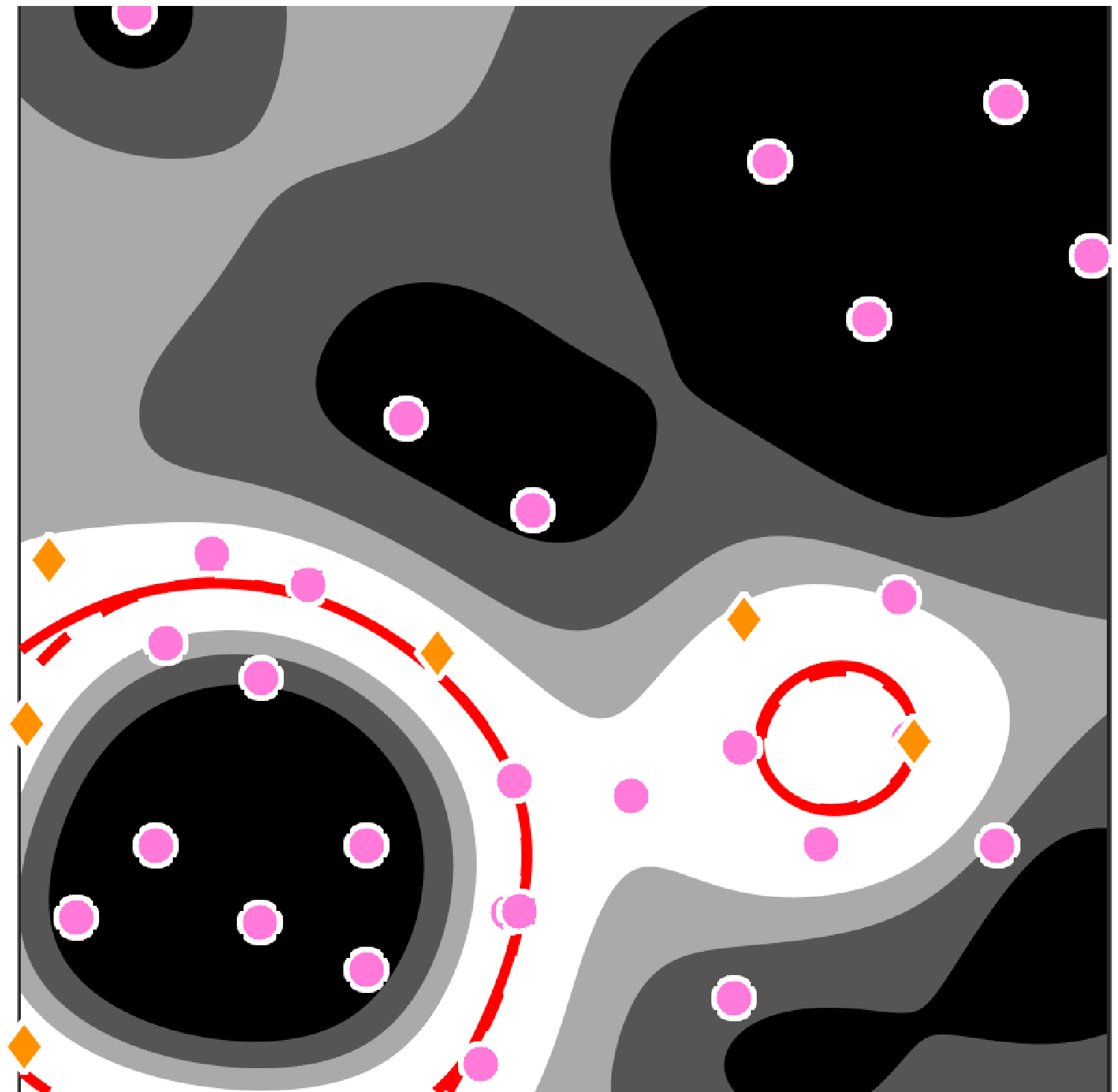}}
\label{subfig:franke_cntr}} \\
\subfigure[NROY identification and sample selection]{
\fboxsep=0mm
\fbox{\includegraphics[keepaspectratio = true, width = .3\linewidth, trim = 75 0 75 1, clip]{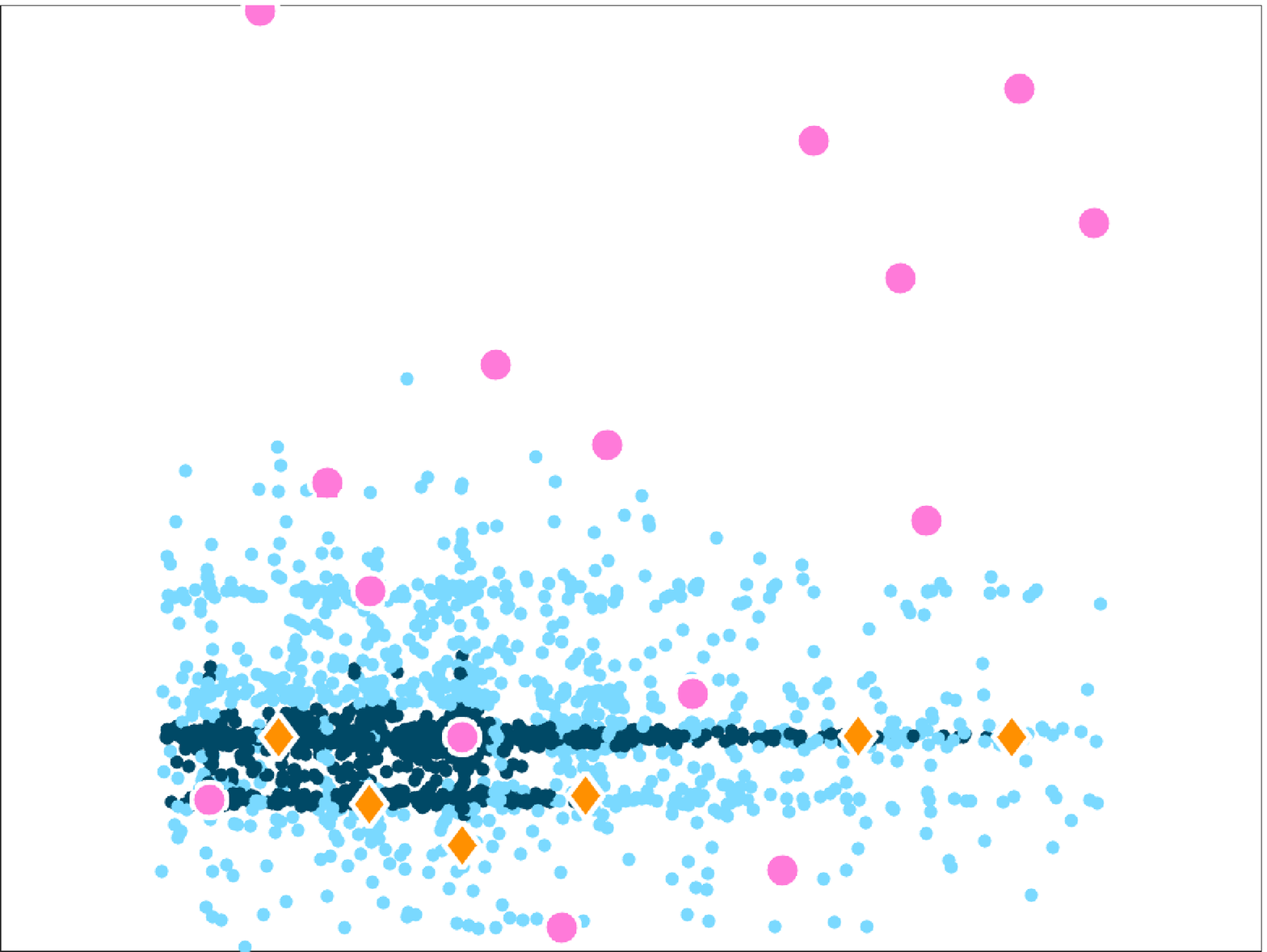}}\qquad
\fbox{\includegraphics[keepaspectratio = true, width = .3\linewidth, trim = 75 0 75 1, clip]{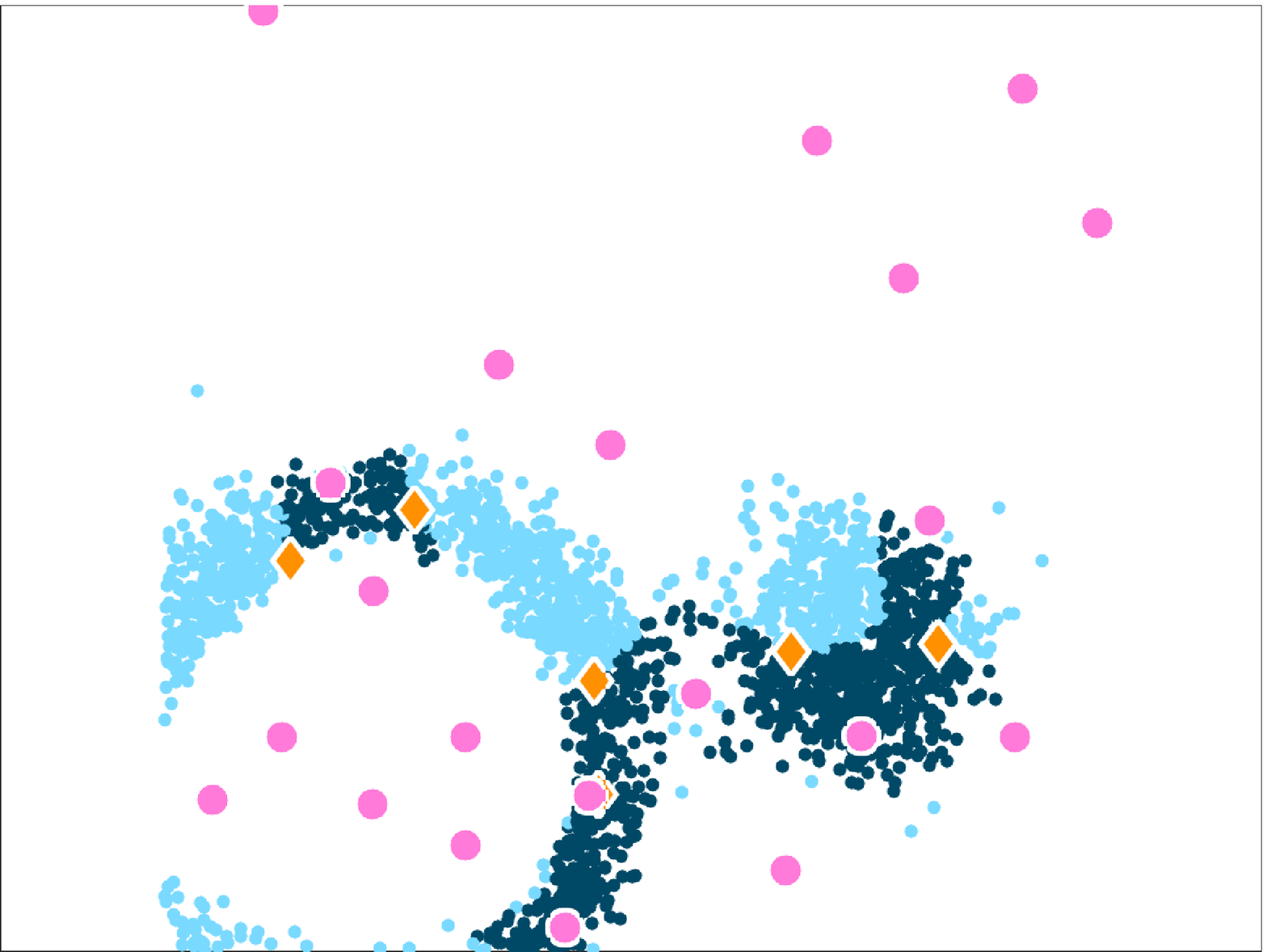}}\qquad
\fbox{\includegraphics[keepaspectratio = true, width = .3\linewidth, trim = 75 0 75 1, clip]{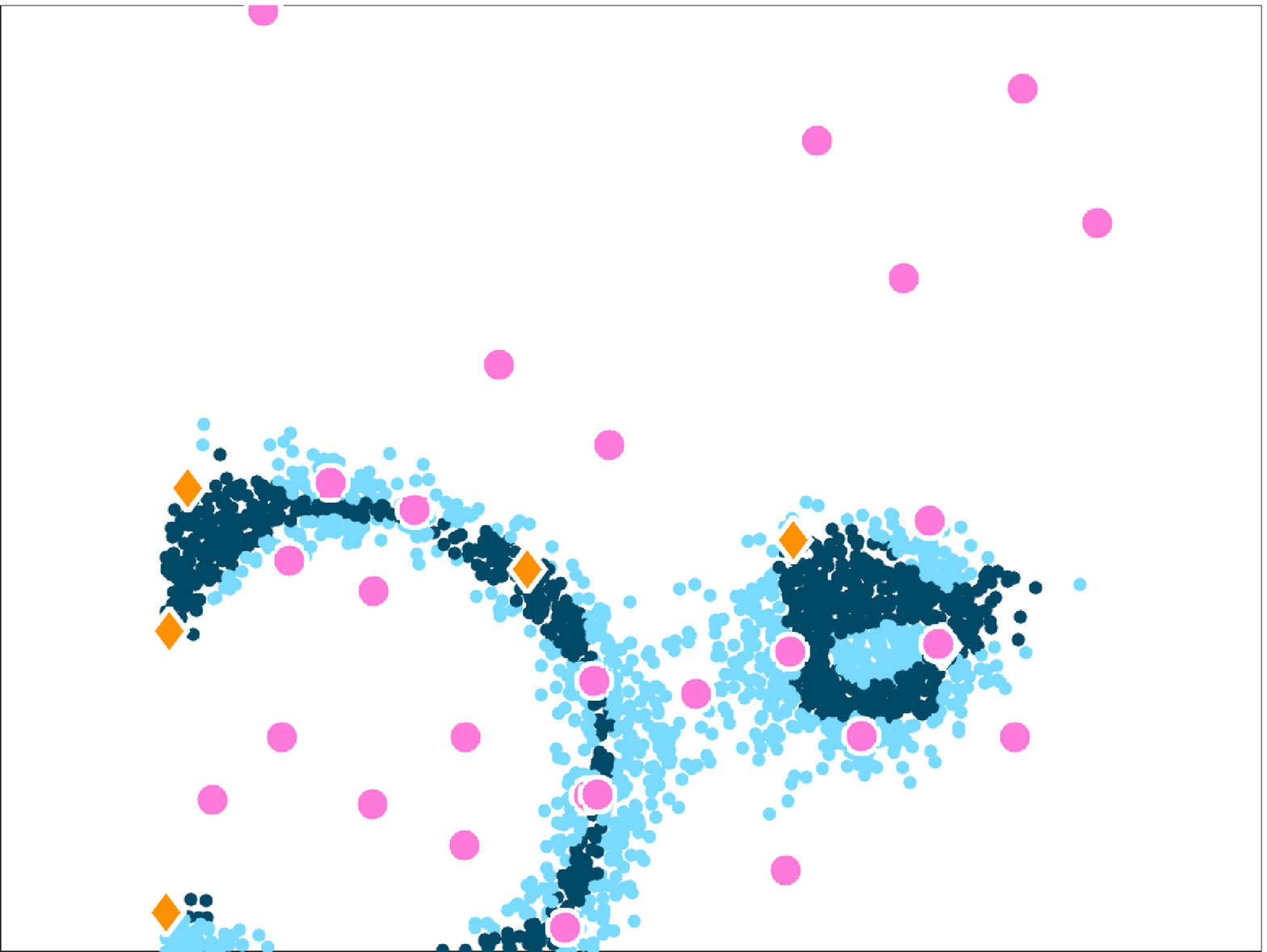}}
\label{subfig:franke_nroy}}

\caption[History matching waves for Franke's function]{\red Results for Franke's
function in the history matching setting. In \cref{subfig:franke_cntr}, contour
levels of the probability of implausibility are shown with lighter shades. Pink
dots represent training runs used for the simulator at wave $t$, and orange
diamonds new points identified in NROY space with good predicted improvement
performance. Each subpanel in \cref{subfig:franke_nroy} shows the samples in
NROY space, with those satisfying a good predicted improvement in darker
colours. The points selected to run the simulator to improve the GP emulator are
shown as orange diamonds. \nc}

\label{fig:franke_samples}

\end{figure}

This experiment uses Franke's function as a simulator. It was first introduced
in the surrogate modeling literature in \citep{Franke1979}. Franke's function is
defined in the two dimensional unit cube, and consists of a sum of three
Gaussian peaks and one smaller dip. The function is defined as

\begin{align}
f(\vec{x}) = \, & \quad \,
0.75 \, \exp \left( - \frac{(9 x_1 - 2)^2}{4} - \frac{(9 x_2 - 2)^2}{4} \right)
+ 0.75 \, \exp \left( - \frac{(9 x_1 + 1)^2}{49} - \frac{9 x_2 + 1}{10} \right)
\nonumber\\
\, & \, + 0.5 \, \exp \left( - \frac{(9 x_1 - 7)^2}{4} - \frac{(9 x_2
- 3)^2}{4} \right) - 0.2\, \exp \left( - (9x_1 - 4)^2 - (9x_2 -7)^2\right).
\end{align}
%
\noindent For the purpose of history matching, the target contour level has been
defined as $z = 0.6$, which results in two disconnected disks, as shown as a
solid red line in \cref{subfig:franke_cntr}. The dashed red lines correspond to
the emulator's predicted contour level of interest. The shaded regions represent
the implausibility contour levels, where lighter colours denote a higher
probability. \red The roughness of the shaded regions is due to the low number
of training runs, which translates as a vague posterior distribution for the GP
hyperparameters.\nc\ As previously discussed, when the available training data
is small, multimodal samplers are able to represent code uncertainty more
robustly \citep{Garbuno2016}.

Panels in \cref{subfig:franke_cntr} show the sequence of waves of the history
matching procedure. The points in blue represent training points that were used
to fit the emulator, whereas orange diamonds depict the chosen points by the
active learning. For the purpose of illustration, the entropic profile discussed
in \cref{sec:active_criteria} was chosen in \cref{fig:franke_samples}. The
samples generated from the NROY space at each wave are shown in
\cref{subfig:franke_nroy}. In this case, the use of sampling algorithms based on
annealed distributions is justified by the complex geometry of the target region
\citep{Zuev2013,Garbuno2016}. In particular, the first panel in
\cref{subfig:franke_nroy} shows that all NROY samples satisfy the property of
being good candidates to improve the emulator. In the same panel, orange dots
denote the chosen points after selecting the top ranking sample shown. For the
remaining subpanels in \cref{subfig:franke_nroy}, light blue dots illustrate
sample points from the NROY space which are not suitable to improve the
emulator. The best candidates are depicted in dark blue following the ranking
from the entropic learning criteria. Also, the space filling interpretation of
the maximin strategy is demonstrated empirically in the first panel (the first
wave of history matching). As noted before, good coverage can be seen in all
panels in \cref{subfig:franke_nroy} by the maximin space filling criteria which
selects the best candidates to improve the emulator.

\begin{figure}[!ht]
\centering
\includegraphics[width = .95\linewidth]{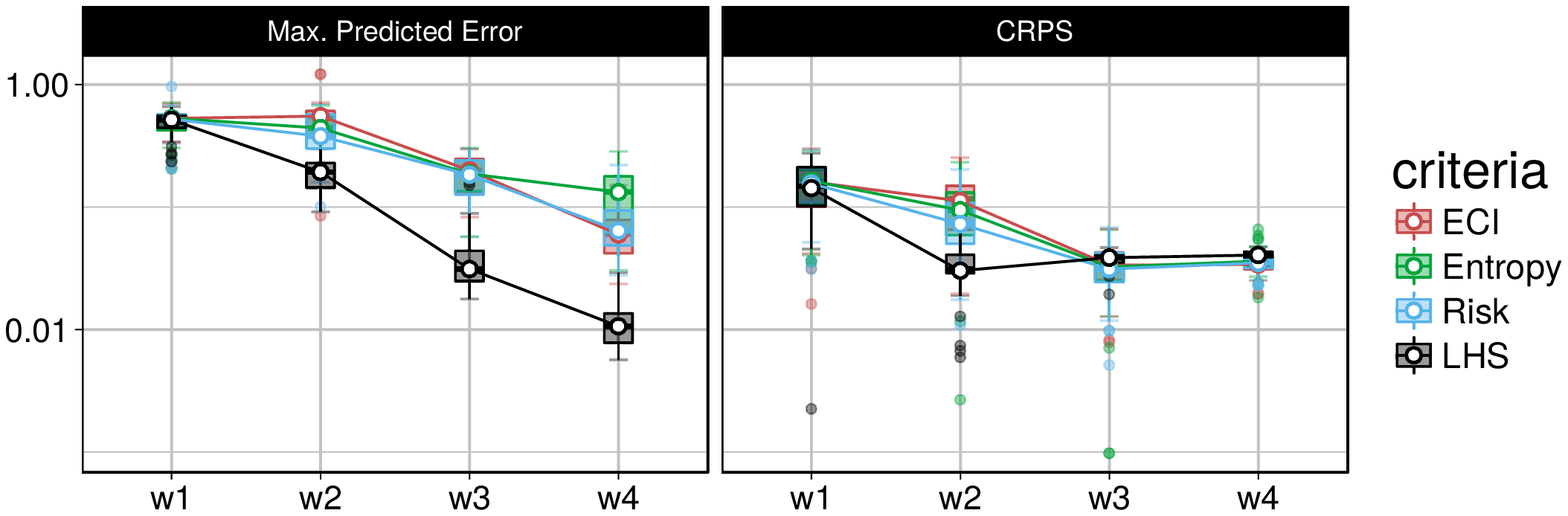}
\caption[History matching summaries for Franke's function]{History matching for
Franke's function. The procedure was performed independently 50 times for each
of the active learning criteria. The results are summarised in boxplots at each
wave. The Maximum predicted error was calculated from the NROY samples at each
wave. Analogously, the reported CRPS was computed as the median from the NROY
samples. Although a space-filling criterion reduces the predicted error further
than the other candidates, the probability statements seem to deteriorate when
compared to the other learning criteria.}
\label{fig:franke_results}
\end{figure}

The procedure was replicated independently 50 times for each learning criteria.
This is depicted \red as boxplots \nc in \cref{fig:franke_results}, where
results are shown for both the maximum predicted error and the CRPS as
iterations advance. \red For the purpose of visualising the trends, the medians
corresponding to each wave are shown connected by a solid line. \nc The
predicted errors steadily decrease for each learning criteria. However, it can
be seen that the decrements in CRPS are marginal in the last wave. The LHS
criterion chooses, among the NROY samples, using the maximin design without any
ranking or prescribed threshold. \red It is important to note that the LHS
sampling scheme appears to decrease the predicted error at each wave. This
strategy seems to be working well as the emulator is improved in terms of
predicting capabilities. However, it should be noted that the other strategies
-- using learning criteria -- are able to improve both the prediction and
variability as measured by the CRPS. \nc

\subsection{The IC fault model}

The following experiment tests the proposed history matching framework in a
physical model. The IC fault model is a cross-sectional simulator of a reservoir
\citep{Tavassoli2005}. Each run is determined by three unknown input parameters,
namely, $h$ (the fault throw), $k_g$ (the good-quality sand permeability) and
$k_h$ (the poor-quality sand permeability). The complexity of the calibration
has made this model become a benchmark to test history matching
\citep{Salter2016}.

The outputs of this simulator are 36-month time series corresponding to three
different properties such as oil production rate, water injection rate and water
production rate. The information of this model is stored as a collection of
159,661 code runs selected uniformly at random in the 3-dimensional cube.
Instead of matching the full time series, only three statistics are chosen as in
\citep{Salter2016}. Those outputs are $o_{24}$ the oil production rate at month
24; $o_{36}$ the oil production rate at month 36; and $w_{36}$, the water
injection rate at month 36.

\begin{figure}[t]
\centering
\includegraphics[width=.95\linewidth]{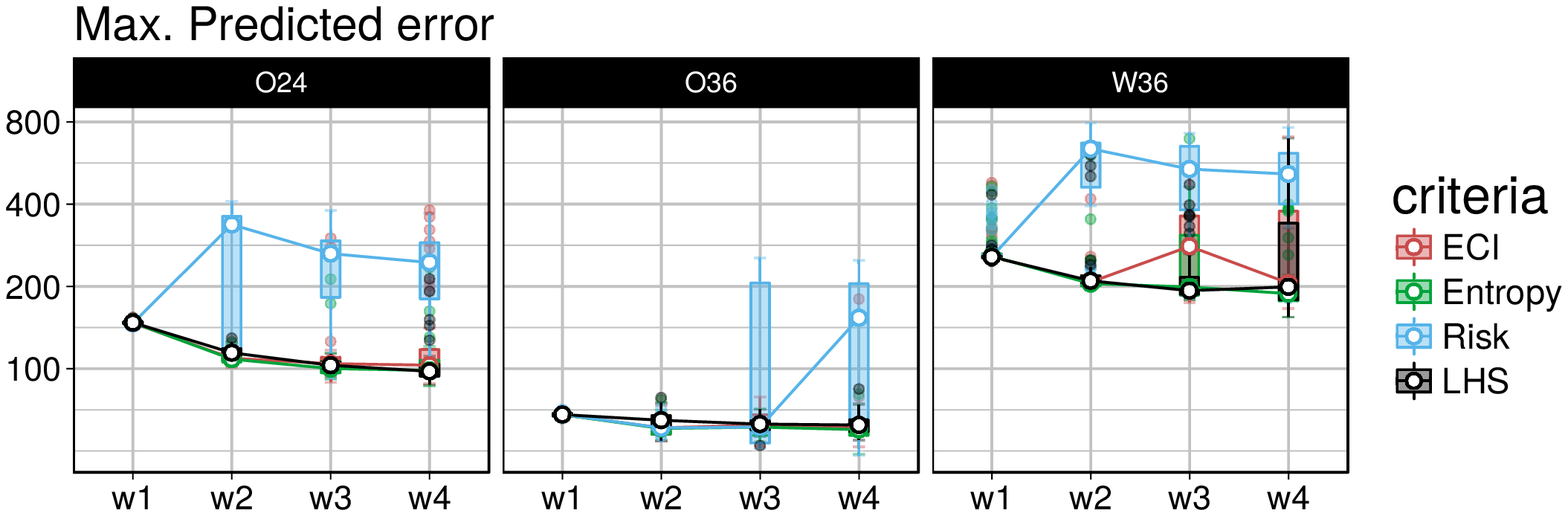}\\ \vspace{-.5cm}
\includegraphics[width=.95\linewidth]{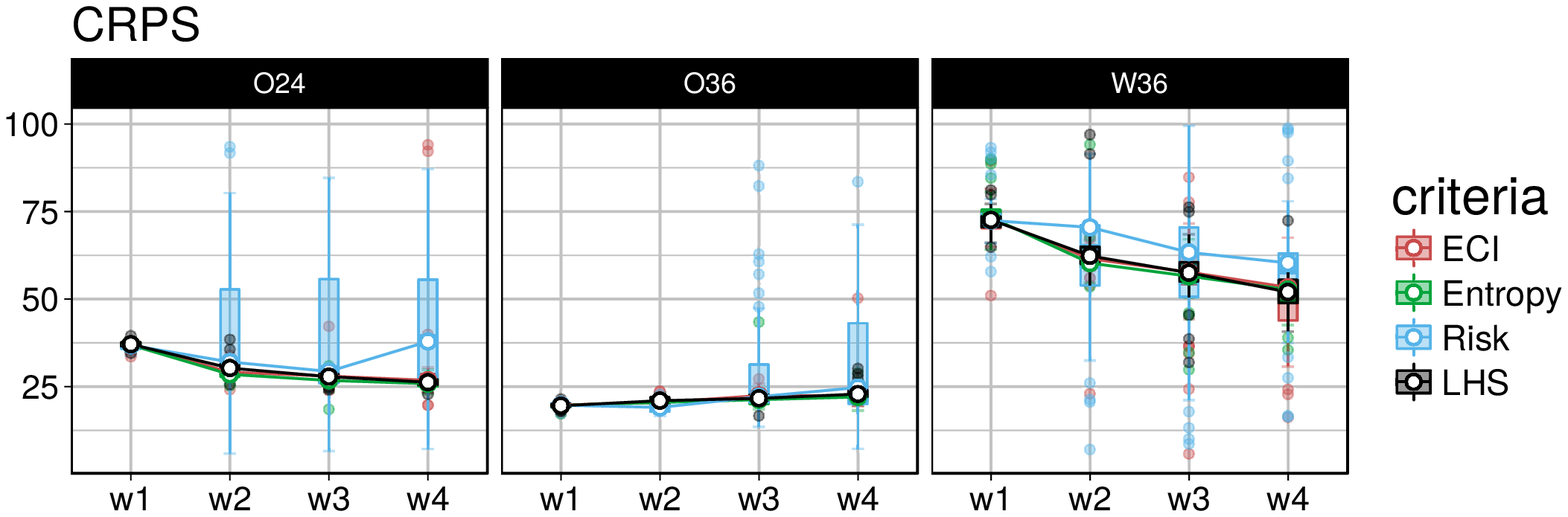}

\caption[History matching summaries for the IC fault model]{Results for the IC
fault model in both performance measurements. Each box corresponds to each
output from left to right. By using the probabilistic Second Maximum
Implausibility measure it can be noted that there is no bias towards an
inaccurate emulator. There is evidence that the oil production rate at month 36,
$o_{36}$, proves difficult to fit with the chosen GP assumptions. Nonetheless,
the history matching procedure is reducing both the uncertainty and the
prediction error for the target contour level. }

\label{fig:icfault_results}
\end{figure}

For experimental purposes, it is assumed that there is no access to such a rich
dataset. In turn, a handful of 60 points are chosen at random from a Latin
hypercube sampling scheme to initialise the procedure. At each wave, an
additional 30 points are selected as discussed in \cref{sec:active_criteria}.
This is done to improve the emulator at the pre-specified target level. Each
output is emulated independently by a GP. In this case, the
implausibility function is a probabilistic version of the Second Maximum
Implausibility Measure of \citep{Vernon2010}, computed from the GP posterior
distribution using a Monte Carlo estimate. This implausibility is used in order
to guard against the possibility that one of the emulators is not performing
accurately. The Second Implausibility Measure is defined as
\begin{align}
I^{(2)}(\x) = \max_i (\, \{ I_{(i)}(\x)\} \setminus I^{(1)}(\x) \,), \label{eq:second_imp}
\end{align}
\noindent where $I_{(i)}$ denotes the implausibility for the $i$-th ouput, and
$I^{(1)}$ denotes the largest Implausibility among all outputs. The target level
to be matched is defined as
\begin{align}
\vec z = (563.6, \, 387.5, \, 917.2 )^\top. \label{eq:icfault_target}
\end{align}
The full history matching procedure was repeated 50 times, each starting with a
different LHS design, and the two performance measures for each wave were
recorded. Results are summarised as boxplots in \cref{fig:icfault_results}. \red
To facilitate the visualisation of the trend, the medians are connected by a
solid line in the same way it was done in \cref{fig:franke_results} . \nc It is
clear that the expected risk learning criteria is both slower and leads to
noisier results for this simulator. In all cases, the oil production rate at
month 36, \ie $o_{36}$, proves too difficult to emulate as seen from the
boxplots in \cref{fig:icfault_results}, which show a slight increase in CRPS and
marginal decrease in predicted error. Nonetheless, history matching overcomes
this limitation and manages to decrease both the expected predicted error and
probabilistic predictions in the target contour level defined in
\cref{eq:icfault_target} for the other outputs. It is important to note that
better results can be achieved if a different GP emulator is trained at every
contour level as in the spirit of \citep{Salter2016}. This work focuses on the
properties of using both the complete probabilistic statements from the Bayesian
posterior of the computer code and the incorporation of active learning criteria
that uses this characterization.

\subsection{Random functions} 

In order to assess the impact of each learning criteria within the proposed
history matching, the following experimental set-up is proposed. It is
inspired by \citep{Hoffman2011}, as it was used to measure the performance of
different acquisition functions used for Bayesian optimisation. The reason to
follow this direction is that there is no generally-agreed set of test
functions for high dimensions in the history matching literature.
The functions to be emulated are generated at random from a GP
prior, as shown in \cref{apx:random_functions}.

Different dimensionalities are chosen in order to understand both the
limitations and strengths of the three active learning criteria for one
dimensional output codes. The LHS discussed in the previous experiments is
included in the comparison. Having different random seeds, it is possible to
replicate the same function for each active learning criteria in each
dimensional setting, thus preserving each set of functions to be compared. In
total, 50 random functions were simulated in each dimensional setting. For each,
the contour level corresponding to the 95\% percentile on the prior seeds as
explained in \cref{apx:random_functions} is chosen as the target for history
matching.

The experiments are chosen in order to include low-dimensional spaces (2D and
3D), medium-sized dimensional spaces (5D and 10D) and large dimensional spaces
(15D and 20D). In the case of larger dimensional settings, dimensionality
reduction techniques can be applied, such as active variable selection
\citep{Vernon2014} or Partial Least Squares \citep{Bouhlel2016}. It is widely
known that GP tend to lose predictive accuracy and robustness with
increasing dimensionality. \red This happens because the kernel used for
the correlation structure relies on some form of Euclidean
distance. Thus, the chosen dimensionalities reflect typical feature spaces where
the GP emulator is able to generalize well. \nc

\begin{figure}[!ht]
\centering
\includegraphics[width= .95\linewidth]{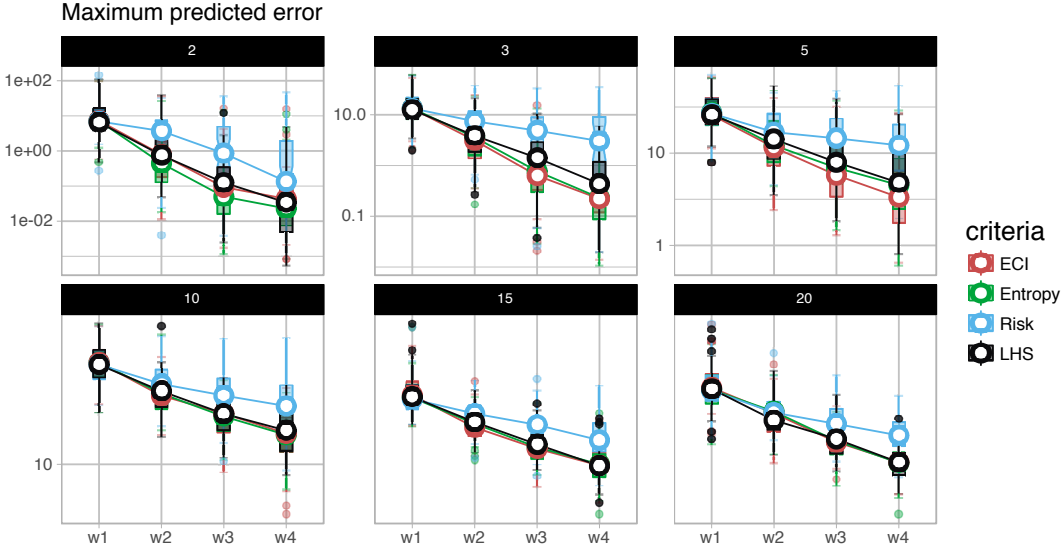}

\caption[Maximum predicted summaries for testbed of functions]{Results for the
maximum predicted error in NROY samples. Each subpanel corresponds to different
dimensional settings indicated in the black headers. The boxplots are generated
by extracting the statistic from each of the replications of the experiment at
each wave. Connecting lines are shown to better appreciate the downward trend as
the iterations succeed. In all dimensional settings the risk learning criteria
is confirmed to be slowest as in \cref{fig:icfault_results}. }

\label{fig:maxerror}
\end{figure}

In the experiments, history matching is not terminated, but the predicted error
is tracked along the iterations. \cref{fig:maxerror} depicts the maximum
predicted error at every wave, for each random function. The results are grouped
in boxplots to show the overall dispersion at each wave. \red Once again, the
medians are connected with lines between waves for the purpose of better trend
visualisation. \nc The space-filling baseline results are shown in black.
Overall, the history matching procedure is successful in reducing the maximum
predicted error. It is important to note that, in any dimensional setting, the
Risk criterion in \cref{eq:risk} shows less improvement as the waves advance.
This is a consequence of the Risk criterion being more susceptible to local
exploration than the other candidates. In low-dimensional settings, the Entropic
profile and the ECI show a slight advantage over the space-filling design. This
is a consequence of a better leverage between the exploitation and exploration
trade-off. In high-dimensional settings, both the Entropic profile and the ECI
show comparable performance to that of the space-filling design. This is
evidence that both criteria are being too general in their rankings and little
exploitation of the surrogate is being used.

\begin{figure}[!ht]
\centering
\includegraphics[width= .95\linewidth]{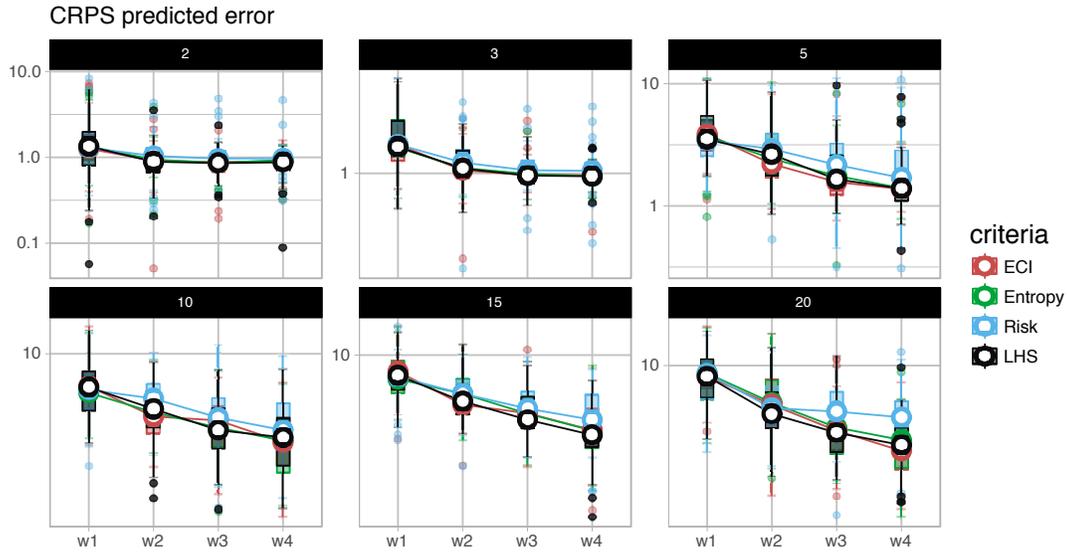}

\caption[CRPS summaries for tested of functions]{Results for the CRPS predicted
error in NROY samples. Each subpanel corresponds to different dimensional
settings indicated in the black headers. The boxplots are generated by
extracting the statistic from each of the replications of the experiment at each
wave. In this case the median of the CRPS is extracted from the NROY samples.
Connecting lines are shown to better appreciate the downward trend as the
iterations succeed. It should be noted that very low dimensional setting the
prediction on the target level do not improves substantially. However, for
larger dimensional spaces it continues to improve.}

\label{fig:crps_results}
\end{figure}

The reduction of the CRPS by the emulation-based history matching is shown in
\cref{fig:crps_results}, again by the trend in the lines that connect the waves
of the procedure. As before, the overall performance is as desired, resulting on
decreasing values of the score. The use of the Risk learning criterion seems to
be hindered again by its lack of willingness to explore the NROY space as the
dimension of the problem increases. As before, the entropic profile and the ECI
show comparable results to a space filling design.

The results show that the Risk learning criterion \cref{eq:risk} is prone to
get trapped in local regions around the best choice. In contrast, for low
and medium dimensional settings, the entropic profile and the ECI show a better
performance in lowering the maximum predicted error (the variance estimated by
the emulator).

For higher dimensional settings, there is no apparent gain in using any of the
criteria discussed above. The use of a space-filling design seems like a safe
choice. However, it should be noted that the learning criteria do achieve a
lower predicted error in low dimensional settings. This should be taken as an
indication that something can be done to enhance the performance on higher
dimensional problems. Recall that the selection of the samples to refine the
emulator is done in two stages. Firstly, the NROY space is identified by an
annealed uniform sampling scheme. Secondly, the samples are ranked accordingly
(choosing a learning criteria) and those that are not able to produce at least a
50\% improvement than the best in the batch are discarded. Following this,
amongst the samples retained, a minimax selection procedure is done, starting by
the top sample. From the results previously exposed, it seems that setting this
50\% target level seems too permissive and that most of the samples are retained
in the procedure. In the end, the minimax selection and a space-filling choice
become equivalent. This is another embodiment of the curse of dimensionality, in
which higher dimensionality requires larger the sampling designs for the
emulator. An alternative, which is subject of current research, is to choose a
batch of good candidates from the learning criteria as it is done in the
Bayesian optimisation setting with the multi-point expected improvement by
\citep{Chevalier2013}.

\section{Conclusions} \label{sec:conclusion}

\red This paper proposes to acknowledge the probabilistic information of the
Gaussian process emulator in history matching applications. This leads to the
incorporation of this probabilistic information into the implausibility
function. \nc The exploitation of this measure is done by sampling with an
annealing schedule as in sequential subset optimisation. This \red sampling
strategy generates \nc uniform samples in the regions defined by a high
probability of being non-implausible. In these regions, the simulator is likely
to replicate the measured data. The ability to sample from complicated
geometries and disconnected regions is achieved by using this form of annealed
sampling. \red These sampling methods \nc have been recently proposed in the
Bayesian inference framework but are flexible enough to accommodate to the
history matching setting \citep{Garbuno2016}. Additionally, the use of active
learning criteria to improve the emulator was also presented. The experimental
results show evidence of better performance when using the expected contour
improvement or the proposed entropic profile, a version adapted to history
matching. This contrasts with random generation of samples by some type of
adapted proposals or rejection-based methods. A family of random functions was
presented to test the effectiveness of this framework, since there is no agreed
collection of history matching test functions. It is important to note that the
learning criteria used in this work could arguably be classified as myopic, in
the sense that the learning functions only take into account the information
available at the current iteration. The extension to finite-horizon criteria, as
in Dynamic programming \citep{Bertsekas2005}, is left as a further research
direction. The learning criteria can potentially decrease the number of samples
to be considered and achieve comparable results to that of using the whole set
of points, as in LHS. Note that theoretical results in \citep{Stuart2018} show
that the GP emulator converges to the true simulator with rates depending on the
coverage of the training point design. The experimental results here suggest
that certain alternatives can achieve similar consistency. Also, the results
shown for the IC-Fault model enhances the need to study further multi-output
history matching application. A direction of current research is the use of more
general learning criteria, or acquisition functions, that mimic batch
optimisation.

\section*{Acknowledgements}

AGI gratefully acknowledges the Consejo Nacional de Ciencia y Tecnolog\'{\i}a
(CONACyT) for the award of a scholarship from the Mexican government for
graduate studies. AGI is supported by the generosity of Eric and Wendy Schmidt
by recommendation of the Schmidt Futures program, by Earthrise Alliance, the
Paul G. Allen Family Foundation, and the National Science Foundation (NSF grant
AGS$-1835860$). FADO acknowledges the support of the Data-centric Engineering
Programme at The Alan Turing Institute, where he was a visiting fellow as part
of the EPSRC grant EP/S001476/1.

\section*{References}
\bibliographystyle{abbrvnat}
\bibliography{histmat}

\appendix

%
%

\section{Expected risk} \label{apx:risk_solution}

The risk criterion is defined by taking into account both the target level $z$
and the predictive probability distribution as learned from the emulator. Note
that the output from a GP at index $x$, denoted as $\fgp(\x)$ is a Gaussian
random variable with mean $m(\x)$ and variance $\sigma^2(\x)$. In the following
the reference to the index $x$ is omitted to ease the exposition.

The risk \red criterion \nc is defined as a piecewise function stemming from two
possibilities. Firstly, as the shortage of reporting $\fgp$ units below the
target level $z$ when in expectation it should have reported a greater quantity.
Secondly, when the report consisted of $\fgp$ units above the target level, when
the expected value was known to be below the target. That is, the risk criterion
evaluated at index $x$ can be written as
\begin{align}
R_z(\x) = \left\lbrace \begin{array}{ll}
\left(\fgp - z\right)_+ & \text{if } m \leq z \\
\left(z - \fgp\right)_+ & \text{if } m > z
\end{array} \right. .
\end{align}
The expected value of the risk is computed following $\fgp \sim \N(m, \sigma^2)$.
The expected risk in the set $m \leq z$ is
\begin{align}
\E [R_z^-] & = \E [ (z - \fgp)_+ ] \\
& = \int_{-\infty}^z (z - \fgp) \, \pi(\fgp) \, \rd \fgp \\
& = \sigma \, \left[ \left( \frac{z - m}{\sigma} \right) \Phi\left( \frac{z - m}{\sigma} \right) + \phi \left( \frac{z - m}{\sigma} \right) \right] \\
& = \sigma \, \left[ \, \bar{z} \, \Phi(\bar{z}) + \phi \left( \bar{z} \right) \right],
\end{align}
where $\pi(\cdot)$ denotes the density for the output of the simulator; $(a)_+ =
\max\{0,a\}$; $\bar{z}$, the standardised target level; and, $\Phi$ and $\phi$
the cumulative and density functions of a standard Gaussian random variable.
\red The expected risk can be computed under an analogous procedure for the
complementary set as \nc
\begin{align}
\E [R_z^+] & = \E [ (\fgp - z)_+ ] \\
& = \int_{z}^{\infty} (\fgp-z) \, \pi(\fgp) \, \rd \fgp \\
& = \sigma \, \left[ \, -\bar{z} \, \Phi( -\bar{z} ) + \phi \left( \bar{z} \right) \right].
\end{align}
\red The expected risk can be computed using the previous results as \nc
\begin{align}
\E[R_z(\x)] = \sigma(\x) \left[ -\text{sign}(\bar{z}) \, \bar{z} \,
\Phi\left(-\text{sign}(\bar{z}\right) \, \bar{z}) + \phi(\bar{z})\right].
\end{align}

\section{Entropic profile} \label{apx:entropy_solution}

\red The entropic profile measures the amount of information the emulator $\fgp$
contains for the interval comprised of $k$ standard deviations around the target
level $z$. Pukelsheim's rule determines an appropriate $k$ under the assumption
of a unimodal distribution for the emulator output to characterise the NROY
space. Thus, the entropic profile of the emulator response is computed as \nc
\begin{align}
H_z(\x) & = \Biggm| - \int_{z - k \sigma}^{z + k \sigma} \log \pi(\fgp) \, \pi(\fgp) \, \rd \fgp\Biggm| \\
   & = \Biggm| - \int_{z - k \sigma}^{z + k \sigma} \left[ - \frac{(\fgp-m)^2}{2\sigma^2} - \log\left(\sqrt{2 \pi \sigma^2} \right) \right] \, \pi(\fgp) \, \rd \fgp\Biggm| \\
   & = \Big| \left[ \ln\left(\sqrt{2\pi \sigma^2}\, \right) + 0.5\right] \,
\left[\Phi(z_2) - \Phi(z_1)\right] - 0.5 \, \left[ z_2
\phi(z_2) - z_1\phi(z_1)\right]\Big|,
\end{align}
where $z_1$ and $z_2$ denote the standardised threshold levels of
the interval. That is, $z_1 = (z-m-k\, \sigma)/\sigma$ and $z_2 =
(z-m+k\, \sigma)/\sigma$. The last inequality is obtained after applying
well-known properties of the integrals of standard Gaussian densities.

\section{Random functions from a GP prior} \label{apx:random_functions}

\red In this paper we propose to generate random test cases from a GP
prior as there are no standard test functions for history matching in increasing
dimensional settings. This is a similar strategy followed in
\citep{Hoffman2011}, who used the posterior mean of a GP as a
random function.

The process is summarised as follows. For each dimensional setting $d$, we
generate $n = 100 \times d$ points uniformly at random from $[0,1]^d$. Let us
denote these chosen seeds as ${\bf X} \in \R^{n \times d}$.\nc\ The lengthscales
of a Mat{\`e}rn kernel, $\boldsymbol \varphi$, are generated from a uniform
random vector in the cube $[0,2]^d$ and the signal noise is chosen as 10. The
random choice of seeds and lengthscales generate different functions during this
process. The choice of the signal noise to be the same for every test case
enables the comparison in terms of predicted error and CRPS among the functions
within the same dimensional setting.

The evaluation of the random function is performed as follows. The Gaussian
process prior defines a multivariate Gaussian distribution for the output on the
seeds $\vec{f} \sim \N(0, 10^2 {\bf K})$, where ${\bf K}$ denotes the covariance
matrix using the Mat{\'e}rn kernel, the seeds ${\bf X}$ and lengthscales chosen
as above. Thus, given a set of training input configurations, say ${\bf X}'$,
the output of the random function is
\begin{align*}
\vec{y} = {\bf k}_*^\top {\bf K}^{-1}\vec{f}
\end{align*}

\noindent where ${\bf k}_*$ is the column vector of pairwise evaluations of the
chosen kernel between each training run ${\bf X}'_i$ and all random seeds ${\bf
X}$.

\end{document}